\documentclass[useAMS,usenatbib]{mn2e}
\usepackage{amsmath, amssymb}
\usepackage{graphicx}
\usepackage{wasysym}

\title[Metallicity, stellar mass and star formation in galaxies]{The relation between metallicity, stellar mass and star formation in galaxies: an analysis of observational and model data}
\author[R. M. Yates, G. Kauffmann \& Q. Guo]{Robert M. Yates$^{1}$\thanks{Email: robyates@mpa-garching.mpg.de}, Guinevere Kauffmann$^{1}$ \& Qi Guo$^{2}$\\
$^{1}$ Max Planck Institut f$\ddot{u}$r Astrophysik, Karl-Schwarzschild-Str. 1, 85741, Garching, Germany\\
$^{2}$ National Astronomical Observatories, Chinese Academy of Sciences, Beijing, 100012, China}
\begin{document}
\date{Accepted ??. Received ??; in original form ??}
\maketitle

\begin{abstract}
We study relations between stellar mass, star formation and gas-phase metallicity in a sample of 177,071 unique emission line galaxies from the SDSS-DR7, as well as in a sample of 43,767 star forming galaxies at $z=0$ from the cosmological semi-analytic model \textsc{L-Galaxies}. We demonstrate that metallicity is dependent on star formation rate at fixed mass, but that the trend is \textit{opposite} for low and for high stellar mass galaxies. Low-mass galaxies that are actively forming stars are more metal-poor than quiescent low-mass galaxies. High-mass galaxies, on the other hand, have lower gas-phase metallicities if their star formation rates are small. Remarkably, the same trends are found for our sample of model galaxies. By examining the evolution of the stellar component, gas and metals as a function of time in these galaxies, we gain some insight into the physical processes that may be responsible for these trends. We find that massive galaxies with low gas-phase metallicities have undergone a gas-rich merger in the past, inducing a starburst which exhausted their cold gas reservoirs and shut down star formation. Thereafter, these galaxies were able to accrete metal-poor gas, but this gas remained at too low a density to form stars efficiently. This led to a gradual dilution in the gas-phase metallicities of these systems over time. These model galaxies are predicted to have lower-than-average gas-to-stellar mass ratios and higher-than-average central black hole masses. We use our observational sample to confirm that real massive galaxies with low gas-phase metallicities also have very massive black holes. We propose that accretion may therefore play a significant role in regulating the gas-phase metallicities of present-day massive galaxies.
\end{abstract}

\begin{keywords}
ISM: abundances -- Galaxies: abundances -- Galaxies: star formation -- Galaxies: fundamental parameters -- Galaxies: evolution
\end{keywords}

\section{Introduction} \label{sec:Introduction}
\LARGE{M}\normalsize etals are ubiquitous throughout galaxies. They are synthesised in stars and liberated into the interstellar medium (ISM) when stars shed their outer gaseous envelopes towards the end of their lives, and in some cases also into the intergalactic medium (IGM) when the highest mass stars explode as supernovae. The amount of metals in the diffuse gas around galaxies also determines the rate at which it is able to cool and form stars. Metallicity is, therefore, one of the key physical properties of galaxies, and understanding the processes that regulate the exchange of metals between stars, cold interstellar gas, and diffuse surrounding gas can help us understand the physical processes that govern galaxy evolution in general.

The metallicity of stars and gas in galaxies is known to correlate strongly with their luminosities, circular velocities and stellar masses (e.g. \citealt{L79,G02,T04,G05}). However, the physical processes that drive these correlations are not yet fully understood.

\citet{MB71} and \citet{L74} first suggested that interstellar gas can be driven out of galaxies by supernova explosions as \textit{galactic outflows}. They predicted that galaxies of smaller mass have lower metal abundances because their lower escape velocities allow freshly enriched gas to be more efficiently removed. 

The input of energy from supernova explosions is now routinely incorporated into hydrodynamical simulations of galaxy formation, either in the form of thermal heating or `kinetic feedback', whereby radial momentum kicks are imparted to particles surrounding sites of star formation in the galaxy. Although these simulations are now able to demonstrate that galactic outflows can yield a good match to the observed mass-metallicity relation (e.g. \citealt{K07,S08}), this does not mean that outflows are the only process at work in regulating the metallicities of real galaxies.

\citet{B07} argue that a \textit{mass-dependent star formation efficiency} (SFE) is required to explain observations, in addition to supernova feedback. In this scenario, less massive galaxies convert their gas reservoirs into stars over longer timescales than more massive galaxies. Therefore, less massive galaxies have higher gas-to-stellar mass ratios and are consequently less metal-rich. Motivated by their work on momentum-driven winds in smoothed particle hydrodynamics (SPH) simulations, \citet{FD08} have also suggested that both metal-rich outflows at all masses and a variable star formation efficiency must play roles in explaining the observed mass-metallicity relation for nearby galaxies.

Alternatively, \citet{D04} suggest that \textit{metal-poor infall} can regulate metallicity in disc galaxies. Given that lower mass galaxies tend to have lower star formation rates (due to the fact that their cold gas densities are lower), a net dilution takes place when the time-scale for star formation falls below the time-scale for accretion of metal-poor gas. This would drive down the low mass end of the mass-metallicity relation.

Finally, \textit{variations in the initial mass function} (IMF) have also been cited as another factor that might influence the mass-metallicity relation. \citet{KWK07} propose that a SFR-dependent (and therefore stellar mass-dependent) IMF causes different galaxies to produce different effective oxygen yields. This hypothesis is based on the premise that most stars form in stellar clusters and that smaller, less actively star-forming galaxies are dominated by clusters with lower masses, containing a smaller fraction of massive, oxygen-producing stars. We note, however, that there is little observational evidence for systematic IMF variations between stellar clusters (e.g. \citealt{El06,FDK11}).

The fact that the interpretation of the mass-metallicity relation is subject to considerable ambiguity has prompted a number of authors to consider alternative ways of quantifying metallicity in galaxies. For example, higher dimensional relations that include additional physical properties could  
provide better constraints on the processes that regulate metallicity. Recently, \citet{M10} have proposed such an extension to the mass-metallicity relation, called the fundamental metallicity relation (FMR). In that work, it was shown that the gas-phase metallicities ($Z$) of both local and high-redshift galaxies were dependent on both stellar mass ($M_{*}$) and star formation rate (SFR). The FMR provides a prediction of the metallicity of local galaxies with a 1$\sigma$ scatter of only $\sim 0.05$ dex. This is a substantial improvement on the mean scatter of $\sim 0.1$ dex reported by \citet{T04} for the $M_{*}$-$Z$ relation. Metallicity was also found to be strongly dependent on SFR at low stellar masses, but only very weakly dependent on SFR at high stellar masses.

\citet{M10} further showed that the FMR describes galaxies out to $z\sim 2.5$. The observed evolution of the $M_{*}$-$Z$ relation was therefore attributed to migration of galaxies along the FMR plane to higher masses and lower star formation rates over time.

Prompted by these findings, in this work we study higher dimensional relations between metallicity and a variety of physical galactic properties. We
examine whether the $M_{*}$-$Z$ relation exhibits additional dependences on SFR, specific SFR (sSFR) and $M_{\textnormal{gas}}/M_{*}$. Comparisons are made between the results from observational data and the predictions from semi-analytic models of galaxy formation implemented within a high resolution simulation of the evolution of dark matter in a `concordance' $\Lambda$CDM cosmology.

In Section \ref{sec:The Samples} we present our observational sample, extracted from the latest Sloan Digital Sky Survey (SDSS) spectroscopic data. In Section \ref{sec:PDMs}, we explain how we obtain estimates of SFR and $Z$ for our sample. In Section \ref{sec:The FMR in observations} we study the dependence of the $M_{*}$-$Z$ relation on SFR, and the dependence of the SFR-$Z$ and sSFR-$Z$ relations on $M_*$. In Section \ref{sec:The FMR in L-Galaxies} we discuss the latest version of the semi-analytic galaxy formation code \textsc{L-Galaxies} \citep{G10}, implemented on the \textsc{Millennium-II} dark matter N-body simulation \citep{BK09}, and describe how we select samples of simulated galaxies for comparison with the observational data. In Section \ref{Model Results} we present these comparisons and show how trends in the relations between stellar mass, star formation rate and metallicity can be understood in terms of the prescriptions for gas accretion, star formation, supernova and AGN feedback that are implemented in the model. In Section \ref{sec:Discussion}, we discuss the viability of different physical mechanisms in regulating the metallicity of galaxies. Finally, in Section \ref{sec:Conclusions}, we summarize our results.

\section{The Observational Sample} \label{sec:The Samples}
The sample of galaxies analysed in this paper is drawn from the Sloan Digital Sky Survey MPA-JHU Data Release 7 catalogue (hereafter, SDSS-DR7)\footnote{available at; \textit{http://www.mpa-garching.mpg.de/SDSS/DR7}}. This catalogue contains $\sim 900,000$ galaxies with available spectra. The sample cuts used here are the same as those adopted by \citet{T04}, who investigated the $M_{*}$-$Z$ relation using galaxies from the SDSS Data Release 2.

First, we remove all duplicate spectra from the catalogue, reducing it by $\sim 3.2$ per cent and leaving 898,302 galaxies. Then, following \citet{T04}, we take only galaxies with reliable spectroscopic redshifts within the range $0.005<z<0.25$. We then remove all galaxies whose fibre-to-total light ratio is less than 0.1. This is defined as the ratio of the flux given by the SDSS fibre magnitude to that given by the SDSS model magnitude, in the r-band. This cut eliminates galaxies with metallicity measurements that are heavily biased towards the inner regions. We have checked that increasing the minimum fibre covering fraction to 0.35 (following recommendations by \citealt{KE08}), or raising the minimum redshift, does not affect any of the main results presented in this paper.
                                              
We also make cuts to the signal-to-noise ratio (SNR) of some of the key emission lines required to estimate metallicity,
ensuring that SNR(H$\alpha$, H$\beta$, [N\textsc{ii}]$\lambda 6584)>5$. Again, we have checked that raising this threshold to $\textnormal{SNR}\geq10$ does not change our main results. Following \citet{T04}, we also make cuts on the accuracy of some additional parameters that were used to estimate stellar masses in their original analysis. All galaxies for which $\sigma(m_{\textnormal{z}})<0.15$ sinh$^{-1}$(mag), $\sigma(\textnormal{H}\delta_{\textnormal{A}})<2.5$\AA, and $\sigma(\textnormal{D}_{n}4000)<0.1$ are removed. This is done purely to achieve consistency with the original sample selection criteria of \citet{T04}. The stellar masses that we use for our current DR7 analysis are derived using $u,g,r,i,z$ SDSS photometry.

Of those galaxies for which SNR([O$\textsc{iii}]\lambda5007)>3$, AGN hosts were removed following the prescription given by \citet{K03c} for defining AGN in the \citet{BPT81} (BPT) diagram:
\begin{equation}\label{eqn:AGN cut}
\textnormal{log}([\textsc{Oiii}]/\textnormal{H}\beta)>0.61/{\textnormal{log}([\textsc{Nii}]/\textnormal{H}\alpha)-0.05}+1.3\;\;.
\end{equation}
For galaxies with SNR([O$\textsc{iii}]\lambda5007)<3$ , only those with $\textnormal{log}([\textnormal{N}\textsc{ii}]\lambda 6584/\textnormal{H}\alpha)<-0.4$ were retained, thus removing low-ionization AGN from the sample. 

Finally, a cut to the derived values of $M_{*}$ and $Z$ was made, based on the `confidence' with which they were estimated from fits to synthetic spectra or HII region models using \textsc{Cloudy} \citep{F98} (see Section \ref{Sample T2}). The 1$\sigma$ spread in the likelihood distribution of the best-fitting model must be less than 0.2 dex in both quantities for the galaxy to remain in the sample.

The number of objects removed by these cuts, including final cleansing (see Sections \ref{sec:Sample T1} and \ref{Sample T2}), is given in Table \ref{tab:cuts2}. Details of our two final samples, when binned by $M_{*}$ and SFR, are given in Table \ref{tab:samples}.

\begin{table}
\centering
\begin{tabular} {c c}
\hline \hline
Sample cut & No. of objects removed\\
\hline
Removing duplicates & 29,250 \\
Redshift cut & 137,087 \\
Fibre-to-total light cut & 34,203 \\
SNR cut & 372,438\\
$\sigma(m_{\textnormal{z}},\textnormal{H}\delta_{\textnormal{A}},\textnormal{D}_{n}4000)$ cut & 16,027 \\
AGN cut & 98,635 \\
$M_{*}$, $Z$ confidence cut & 62,841 \\
T1 cleansing & 56,580 \\
T2 cleansing & 64,264 \\
\hline \hline
\end{tabular}
\caption{The number of objects removed by each sample cut, in the order they were applied. The final number of objects in Sample T1 and T2 is 120,491 and 112,797, respectively. The difference in final size between the two samples is due to the different requirements for estimating SFR and $Z$ (see text). 93,971 galaxies are common to both samples.}
\label{tab:cuts2}
\end{table}

\section{Estimation of Metallicity and Star Formation Rate} \label{sec:PDMs}
After creating our base sample of 177,071 emission line galaxies, we implement two different procedures for estimating metallicity and star formation rate. 

The first procedure aims to replicate the methods outlined in \citet{M10}, and uses the H$\alpha$-based method of \citet{K98} for deriving SFR, and two of the strong line ratio calibrations described by \citet{M08} for deriving metallicity. We will refer to the list of metallicities, stellar masses and star formation rates generated using this procedure as Sample T1.

The second procedure uses the values for SFR and $Z$ provided by the SDSS-DR7 online catalogue. These estimates utilize emission line fluxes and Bayesian techniques implemented in a library of model galaxies that were created using a combination of stellar population synthesis models from \citet{BC03} and HII region models from \citet{CL01}. The Bayesian estimation technique provides a way of deriving robust errors for each of the derived physical parameters. We will refer to this list of estimates as Sample T2. Comparison of the two procedures provides insight into how sensitive the derived relations between $M_{*}$, SFR and $Z$ are to uncertainties in our SFR and metallicity estimates.  

We note that oxygen abundance is used as a proxy for global gas-phase metallicity throughout this work. We express metallicity in terms of the number of oxygen atoms to hydrogen atoms in the gas component of a galaxy, normalised to the dimensionless quantity $Z = 12+\textnormal{log(O/H)}$. The current determination of the solar oxygen abundance in these units is $Z_{\textnormal{\astrosun}} = 8.69$ \citep{AP01,A09}.

\begin{table}
\centering
\begin{tabular} {c c c c}
\hline \hline
\textbf{Sample T1} & Minimum & Median & Maximum \\
\hline
log($M_{*}$) & 8.925 & 10.125 & 11.175 \\
log(SFR) & -2.025 & -0.525 & 0.975 \\
$12+\textnormal{log(O/H)}$ & 8.72 & 9.01 & 9.14 \\
Redshift & 0.030 & 0.077 & 0.179 \\
\hline
No. of galaxies & 120,491 & & \\
\hline \hline
\textbf{Sample T2} & Minimum & Median & Maximum \\
\hline
log($M_{*}$) & 8.625 & 9.975 & 11.175 \\
log(SFR) & -1.125 & 0.075 & 1.575 \\
$12+\textnormal{log(O/H)}$ & 8.48 & 8.88 & 9.13 \\
Redshift & 0.019 & 0.068 & 0.174 \\
\hline
No. of galaxies & 112,797 & & \\
\hline \hline
\end{tabular}
\caption{Details of the parameter space coverage of our two observational samples. These values are given for data binned by $M_{*}$ and SFR, considering only bins containing $\geq 50$ galaxies.}
\label{tab:samples}
\end{table}

\subsection{Sample T1} \label{sec:Sample T1}
Following \citet{M10}, total stellar masses are taken directly from the SDSS-DR7 catalogue, and corrected from a Kroupa to a Chabrier IMF by dividing by a factor of 1.06. Cleansing the sample of galaxies with: a) uncertain estimates of $M_{*}$ in the catalogue, b) emission lines that have too low signal-to-noise (SNR) for accurate $Z$ estimates, c) estimates of $Z$ from the two diagnostics used that differ by more than 0.25 dex (see below),
reduces the sample to a total of 120,491 galaxies.

SFRs were measured from the H$\alpha$ emission line flux, corrected for dust extinction. The corrected H$\alpha$ flux from each galaxy was converted to luminosity using $L=F_{\textnormal{c}}\times 4\pi D^{2}\times 10^{-17}$, where the distance $D$ to the source was derived from the galaxy's redshift, and $10^{-17}$ is the factor used to normalise SDSS fluxes to units of ergs s$^{-1}$ cm$^{-2}$. Star formation rate was then determined using the fixed conversion from \citet{K98},
%$\textnormal{\textit{ergs}}\;s^{-1}cm^{-2}$

\begin{equation} \label{eqn:L-to-SFR}
\textnormal{SFR}=7.9\times 10^{-42}\; L(\textnormal{H}\alpha)\: [\textnormal{ergs s}^{-1}]\;\;,
\end{equation}
which assumes a case B recombination, an electron temperature of $10^{4}\:$K and a Salpeter IMF. This value was then also corrected to a Chabrier IMF by dividing by 1.7.

The observed H$\alpha$ flux is corrected for external dust extinction using

\begin{equation} \label{eqn:flux-correction}
F_{\textnormal{c}}(\textnormal{H}\alpha)=F_{\textnormal{obs}}(\textnormal{H}\alpha)\;e^{+\tau_{\textnormal{H}\alpha}}\;\;,
\end{equation}
where the H$\alpha$ optical depth $\tau_{\textnormal{H}\alpha}$ can be determined using the wavelength-independent relation $A_{\lambda}=1.086\tau_{\lambda}$ \citep{C94}. An estimate of $A_{\textnormal{H}\alpha}$ can be obtained from the Balmer decrement $B=F_{\textnormal{obs}}(\textnormal{H}\alpha)/F_{\textnormal{obs}}(\textnormal{H}\beta)$ and the following equation:

\begin{equation}\label{eqn:extinction}
A_{\textnormal{H}\alpha}=-2.5\:\textnormal{log}\left(\frac{B}{2.86}\right)\cdot \frac{k_{\textnormal{H}\alpha}}{k_{\textnormal{H}\alpha}-k_{\textnormal{H}\beta}}\;\;,
\end{equation}
where 2.86 is the intrinsic Balmer decrement for a case B recombination with electron density $n_{e}=100 \textnormal{cm}^{-3}$ and electron temperature $T_{e}=10^{4}$K \citep{O89}. These values are roughly appropriate for local star forming galaxies \citep{Ke01,I06,L08,PM11}. The k-values used are $k_{\textnormal{H}\alpha}=2.468$ and $k_{\textnormal{H}\beta}=3.631$ \citep{C01}, and represent the ratio of the extinction at a given wavelength to the intrinsic colour excess in the b-band relative to the v-band ($k_{\lambda}=A_{\lambda}/E(B-V)_{i}$) \citep{C94}.

It should be noted that these SFRs pertain to the region of the galaxy falling within the 3'' diameter SDSS fibre aperture, which probes the inner $\sim$1-9 kpc of the galaxies in our samples. This method was used for Sample T1 in order to follow the procedure adopted by \citet{M10} as closely as possible. The method used in Sample T2 instead estimates the total-SFR using the SDSS photometry to correct for the missing star formation in the outer regions of the galaxy (see Section \ref{Sample T2}).

Metallicity was calculated for Sample T1 using two of the strong line diagnostics calibrated by \citet{M08}. Such diagnostics are often used when direct measurements of the electron temperature $T_{e}$ in the \textsc{Hii} regions of a galaxy are not possible, or at high metallicities ($Z \gtrsim 8.35$), where they no longer provide an accurate estimate of $Z$ due to temperature fluctuations within individual H\textsc{ii} regions and across the whole galaxy \citep{St05,M08} (this is the case for essentially all the galaxies in our sample). In such cases, either an empirical method, utilising other galaxies with measured $T_{e}$ metallicities, or a theoretical method, utilising purely theoretical photoionisation models, can be used \citep{KE08}.

\citet{M08} derived their strong line diagnostics using a combination of empirical and theoretical methods. 259 local galaxies of $Z<8.35$, with $T_{e}$ derived metallicities compiled by \citet{N06} were used, combined with 22,482 SDSS-DR4 galaxies of $Z>8.4$, with metallicities derived using the photoionisation model outlined in \citet{KD02}. The resulting combined calibrations are given by eqn. 1 and table 4 in the \citet{M08} paper.

For our Sample T1, we follow \citet{M10} by taking the average of the metallicities given by the [N\textsc{ii}]$\lambda6584$/H$\alpha$ calibration and the $R_{23}$ calibration as the final metallicity estimate for each galaxy.\footnote{A modified version of the R$_{23}$ ratio was used for Sample T1 which does not require the [O\textsc{iii}]$\lambda$4959 line: $\textnormal{R}_{23}=([\textnormal{O}\textsc{ii}]\lambda3727+3.1\cdot [\textnormal{O}\textsc{iii}]\lambda5007)/\textnormal{H}\beta$. This version provides a more robust diagnostic for metallicity when the SNR of the oxygen lines is low (F. Mannucci, private communication).} We also corrected all line fluxes for dust, following \citet{C89}. This lowers the metallicities estimated via R$_{23}$ by $\sim0.03$ dex at the highest masses, but makes very little difference to those estimated via [N\textsc{ii}]$\lambda6584$/H$\alpha$ because the two lines involved are of very similar wavelengths.

\subsection{Sample T2} \label{Sample T2}
Our second sample, Sample T2, utilises the stellar masses, SFRs and metallicities given in the online SDSS MPA-JHU DR7 catalogue. Values have been nominally corrected from a Kroupa to a Chabrier IMF to match Sample T1. \textit{Total} stellar masses and star formation rates are used, whereas metallicities are calculated using emission line fluxes that fall within the spectral fibre. Cleansing the sample of galaxies without robust estimates for $M_{*}$, SFR or $Z$ from the catalogue reduced the sample to 112,797 galaxies. We note that the stellar masses in the SDSS-DR7 catalogue are based on fits to photometric data rather than to Lick indices, as in \citet{K03a}. 

Total-SFRs are calculated using the technique described by \citet{B04}, with improvements to the aperture corrections as detailed by \citet{S07}. 
This method for estimating star formation rates is based on fitting to a grid of photoionisation models derived from the \textsc{Cloudy} code \citep{F98}, as detailed by \citet{CL01}. It thus accounts for the fact that the H$\alpha$-to-SFR conversion factor will depend on metallicity, and it also allows the dust-free value of the Balmer decrement to differ from the `standard' case B value (see \citealt{B04} for a more extensive discussion). The use of total- rather than fibre-SFRs is the main reason why the median star formation rate of galaxies in Sample T2 is 0.6 dex higher than it is in Sample T1 (see Table 2). The most significant shift is seen at low masses, as these galaxies tend to have low redshifts and larger apparent sizes, and are therefore more extended with respect to the fibre aperture than more distant galaxies.

Metallicities are calculated using the same grid of photoionisation models, by finding the model that best matches the observed fluxes of the most prominent optical emission lines ([O\textsc{ii}], H$\beta$, [O\textsc{iii}], H$\alpha$, [N\textsc{ii}] and [S\textsc{ii}]). We refer the reader to 
Appendix A for a brief discussion on the merits of using this Bayesian technique over simpler emission line ratios when studying local, high-$Z$, star-forming samples.

\section{Observational Results} \label{sec:The FMR in observations}
In this section, we examine whether the $M_{*}$-$Z$ relation exhibits additional dependences on SFR. The region of the parameter space of interest for our analysis is $8.6<\textnormal{log(}M_{*}\textnormal{)}<11.2$, $-2.0<\textnormal{log(SFR)}<1.6$ and $8.5<Z<9.2$, as this covers 98 per cent of the galaxies in both our observational samples.

\begin{figure}
\centering
\includegraphics[totalheight=0.3\textheight, width=0.46\textwidth]{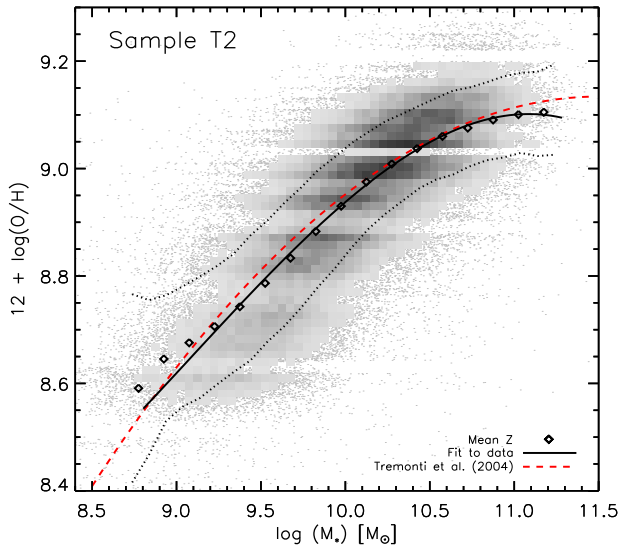}
\caption{The $M_{*}$-$Z$ relation for Sample T2. The galaxy distribution is shown in grey, and a fit to this data is given (black solid line). The mean metallicity in bins of 0.15 dex in stellar mass is plotted as black diamonds. The 1$\sigma$ dispersion in $Z$ about the mean is shown as dotted lines. Also shown is the \citet{T04} fit to data drawn from the SDSS-DR2 (dashed red line).}
\label{fig:T_overallM-Z}
\end{figure}

\begin{figure}
\centering
\includegraphics[totalheight=0.2\textheight, width=0.3\textwidth]{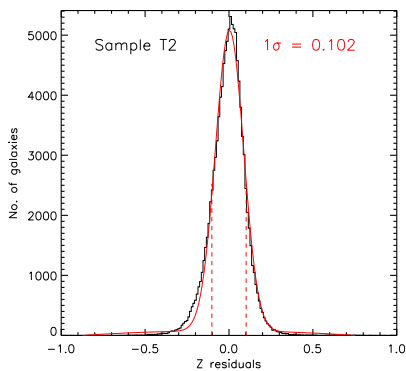}
\caption{The distribution of residuals about $Z$ for the $M_{*}$-$Z$ relation shown in Fig. \ref{fig:T_overallM-Z}. The mean dispersion of $\sim0.10$ dex is the same as that reported by \citet{T04} for their SDSS-DR2 sample.}
\label{fig:T2_residuals}
\end{figure}

\subsection{The $M_{*}$-$Z$ relation} \label{sec:The M-Z relation}
Firstly, we introduce the basic $M_{*}$-$Z$ relation for Sample T2, before analysing any SFR-dependence. In this form, the relation can be seen as an `update' to that of \citet{T04}, who analysed a sample containing half as many galaxies. The $M_{*}$-$Z$ relation for Sample T2 is shown in Fig. \ref{fig:T_overallM-Z}. Individual galaxies are shown in grey, with a 3rd order polynomial fit to the whole population shown as a solid black line, and given by the following equation:

\begin{equation} \label{eqn:T_bestfit}
Z = 26.6864-6.63995x+0.768653x^{2}-0.0282147x^{3}\;\;,
\end{equation}
where $Z = 12+\textnormal{log(O/H)}$ and $x = \textnormal{log(}M_{*}/\textnormal{M}_{\textnormal{\astrosun}}\textnormal{)}$. A fit to the \citet{T04} mass-metallicity relation (red dashed line) is plotted for comparison. The standard deviation about the best fit from residuals is 0.102 dex, as shown by Fig. \ref{fig:T2_residuals}. This is the same as the dispersion calculated for the original \citet{T04} sample. 

Our new fit indicates a somewhat more linear relation between mass and metallicity below $\sim10^{10}\textnormal{M}_{\textnormal{\astrosun}}$. Above this mass, the relation flattens and there is a hint of a turnover at the very highest masses. Turnovers in the $M_{*}$-$Z$ relation at high mass are not uncommon in the literature \citep{Z94,KD02,KK04,M91,PP04}, however they have not been widely discussed in terms of their physical significance or with reference to star formation rates.

\subsection{The $M_{*}$-$Z$ relation, as a function of SFR} \label{sec:The M-Z relation, as a function of SFR}
\begin{figure*}
\centering
\begin{tabular}{c c}
\includegraphics[totalheight=0.3\textheight, width=0.46\textwidth]{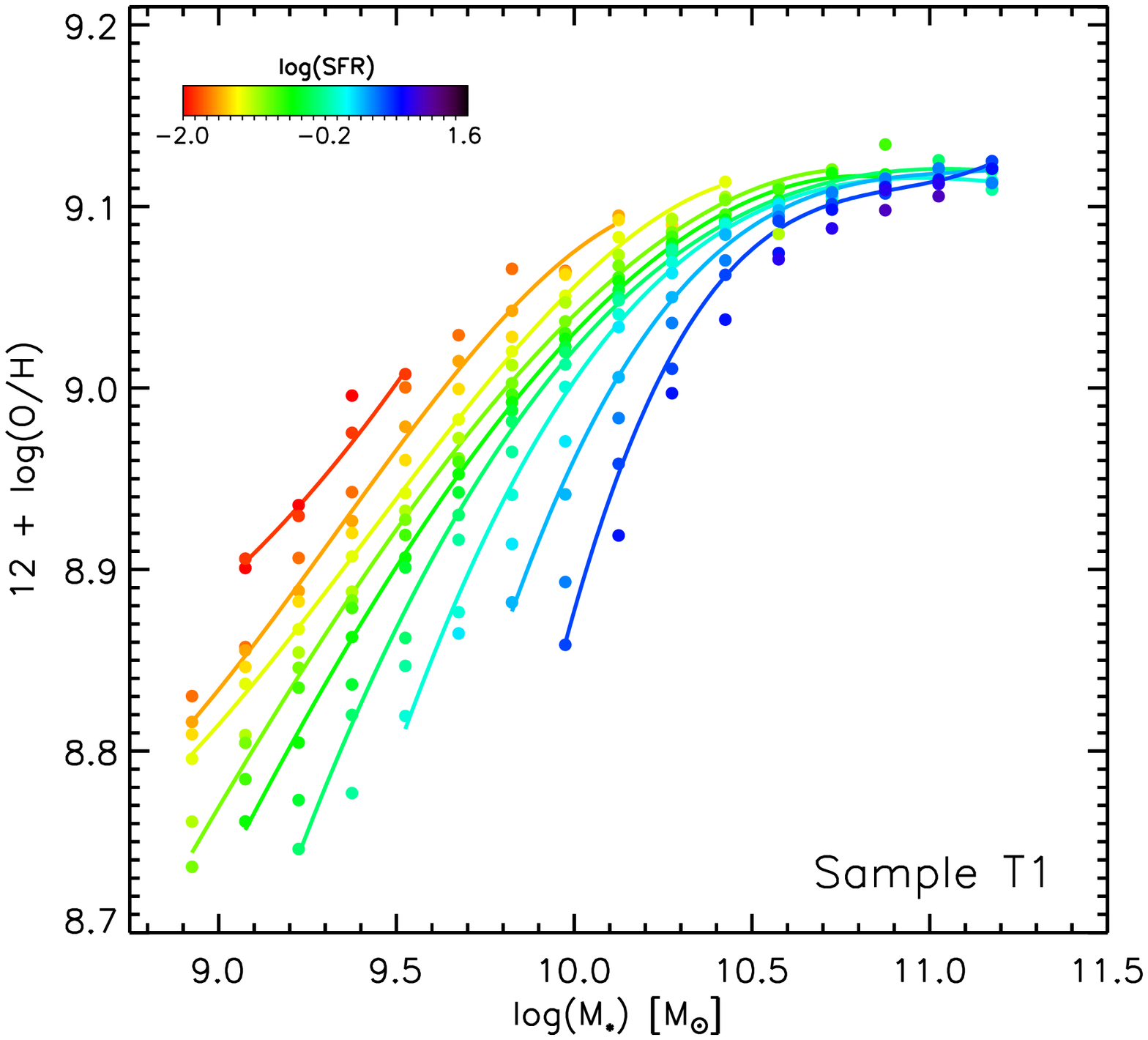} &
\includegraphics[totalheight=0.3\textheight, width=0.46\textwidth]{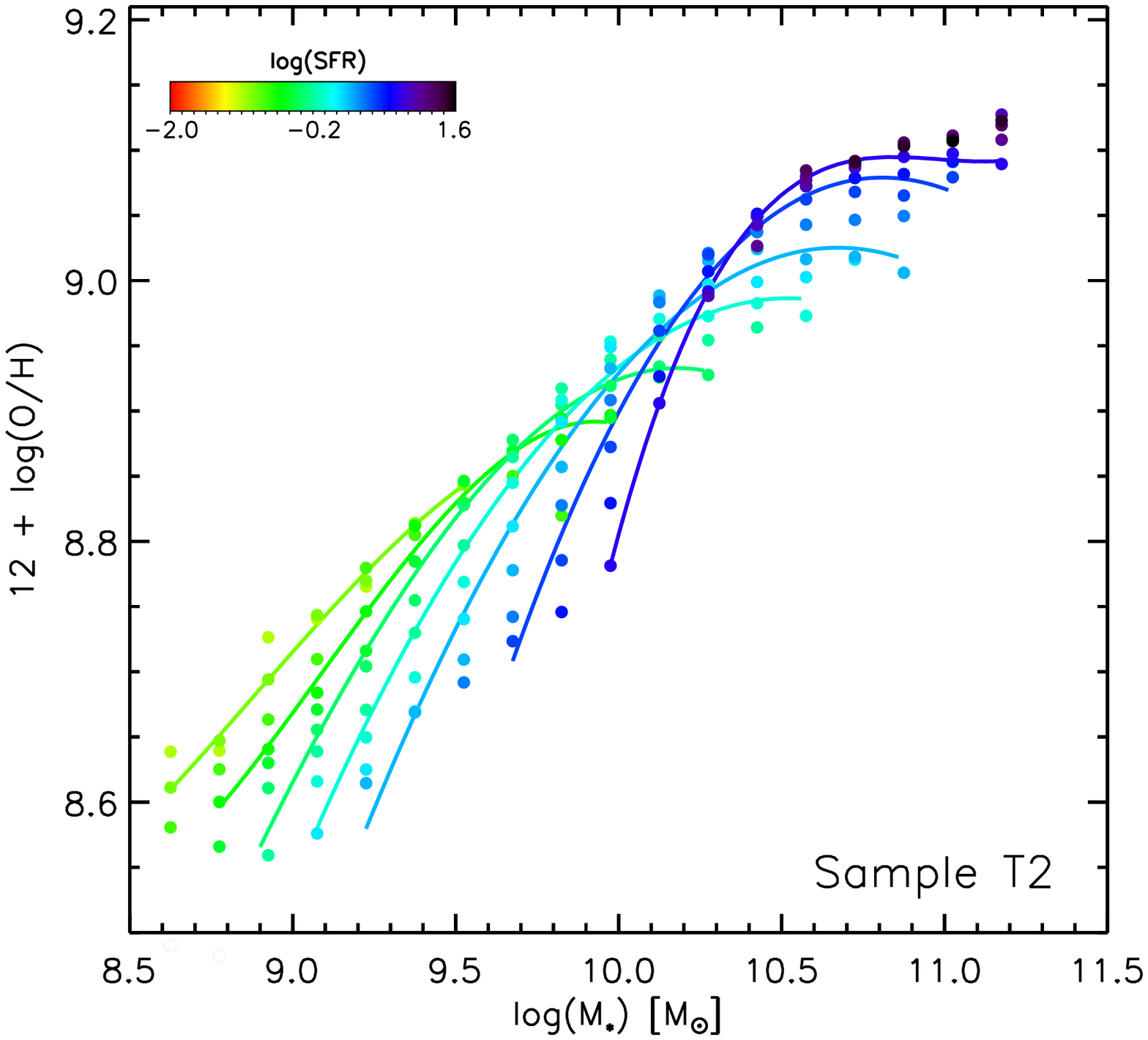} \\
\end{tabular}
\caption{The $M_{*}$-$Z$ relation for Sample T1 (left panel) and Sample T2 (right panel). Filled circles show the mean metallicities of galaxies binned by $M_{*}$ and SFR, for bins containing $\geq 50$ galaxies. Binned data is coloured by SFR, as are the fits at fixed SFRs (solid lines), plotted for log(SFR) = -1.875, -1.575, -1.275, -0.975, -0.675, -0.375, -0.075, 0.225, 0.525 for Sample T1, and log(SFR) =  -0.975, -0.675, -0.375, -0.075, 0.225, 0.525, 0.825 for Sample T2. The dependence of $Z$ on SFR for the two samples is clearly different. For example, Sample T2 exhibits a clear dependence at high-mass.}
\label{fig:M+T_M-Z_mk3}
\end{figure*}

In order to study the dependence of the $M_{*}$-$Z$ relation on SFR, we follow a similar approach to \citet{M10}, binning galaxies by $M_{*}$ and SFR, and calculating the mean metallicity in each bin. Bins are of width 0.15 dex in both dimensions, and only those which contain $\geq 50$ galaxies are plotted. Fig. \ref{fig:M+T_M-Z_mk3} shows this $M_{*}$-$Z$ relation for Sample T1 (left panel) and Sample T2 (right panel). The data are coloured by star formation rate, as are fits to the relation at a series of fixed SFRs (solid lines).

The left panel of Fig. \ref{fig:M+T_M-Z_mk3} clearly shows the result previously described by \citet{M10}: there is an increase in metallicity with increasing mass, but also a clear and ordered dependence of metallicity on SFR at fixed mass. Metallicity depends strongly on SFR at low masses, but is virtually independent of SFR at high masses in this sample. The 1$\sigma$ spread about the median $Z$ for Sample T1 is 0.07 dex, compared to 0.08 dex reported by \citet{M10}.

The $M_{*}$-$Z$ relation for Sample T2 is somewhat different. It extends down to lower stellar masses and metallicities, and has a higher median star formation rate. Additionally, rather than an ordered segregation of the relation for fixed SFR, Sample T2 exhibits a `twist', where low-SFR galaxies change from being the most metal-rich at low masses, to the least metal-rich at high masses. Sample T2, therefore, exhibits a SFR-dependence at both low and at high mass, with a broad transition region between the two regimes around $M_{*} \sim 10^{10.2}\textnormal{M}_{\textnormal{\astrosun}}$. We note that we obtain the same features for our observational samples when using the sample selection criteria described by \citet{M10}, rather than those outlined in Section \ref{sec:The Samples}. This indicates that the choice of sample cuts does not significantly affect our findings.

The extension to lower stellar masses seen for Sample T2 is mainly due to the removal of galaxies from Sample T1 during cleansing. Low-redshift galaxies ($z<0.03$) are excluded from Sample T1 because the [O\textsc{ii}]$\lambda3727$ line (required for estimating metallicity using the R$_{23}$ ratio) is not measurable. These galaxies can remain in Sample T2, extending the $M_{*}$-$Z$ relation down a further $\sim 0.3$ dex in stellar mass.

The shift in SFR seen in Sample T2 is  explained by a combination of factors. First, total- rather than fibre-SFRs are used (see Section \ref{Sample T2}). Second, the simpler H$\alpha$-based method used for Sample T1 yields lower SFR estimates for high-$M_{*}$ galaxies of given H$\alpha$ luminosity. This is because the conversion factor in Eqn. \ref{eqn:L-to-SFR} is dependent on metallicity, and hence stellar mass. \citet{B04} report that this Kennicutt value is most accurate for galaxies with stellar masses $\sim 10^{9.5} \textnormal{M}_{\textnormal{\astrosun}}$, and will underestimate the SFR in more massive galaxies. Note, however,  that the intrinsic Balmer decrement for case B recombination -- the case B ratio -- is also metallicity- and mass-dependent. A fixed case B ratio can over-estimate the attenuation due to dust by up to $\sim 0.5$ mag for the most massive, metal-rich galaxies \citep{B04}. This will counteract somewhat the underestimate in SFR from a fixed conversion factor for higher mass galaxies. None the less, the use of fibre-SFRs and a fixed conversion factor combine to shift the median value of SFR down from log(SFR) = 0.075 $\textnormal{M}_{\textnormal{\astrosun}}/$yr for Sample T2 to log(SFR) = -0.525 $\textnormal{M}_{\textnormal{\astrosun}}/$yr for Sample T1.

\begin{figure}
\centering
\includegraphics[totalheight=0.2\textheight, width=0.3\textwidth]{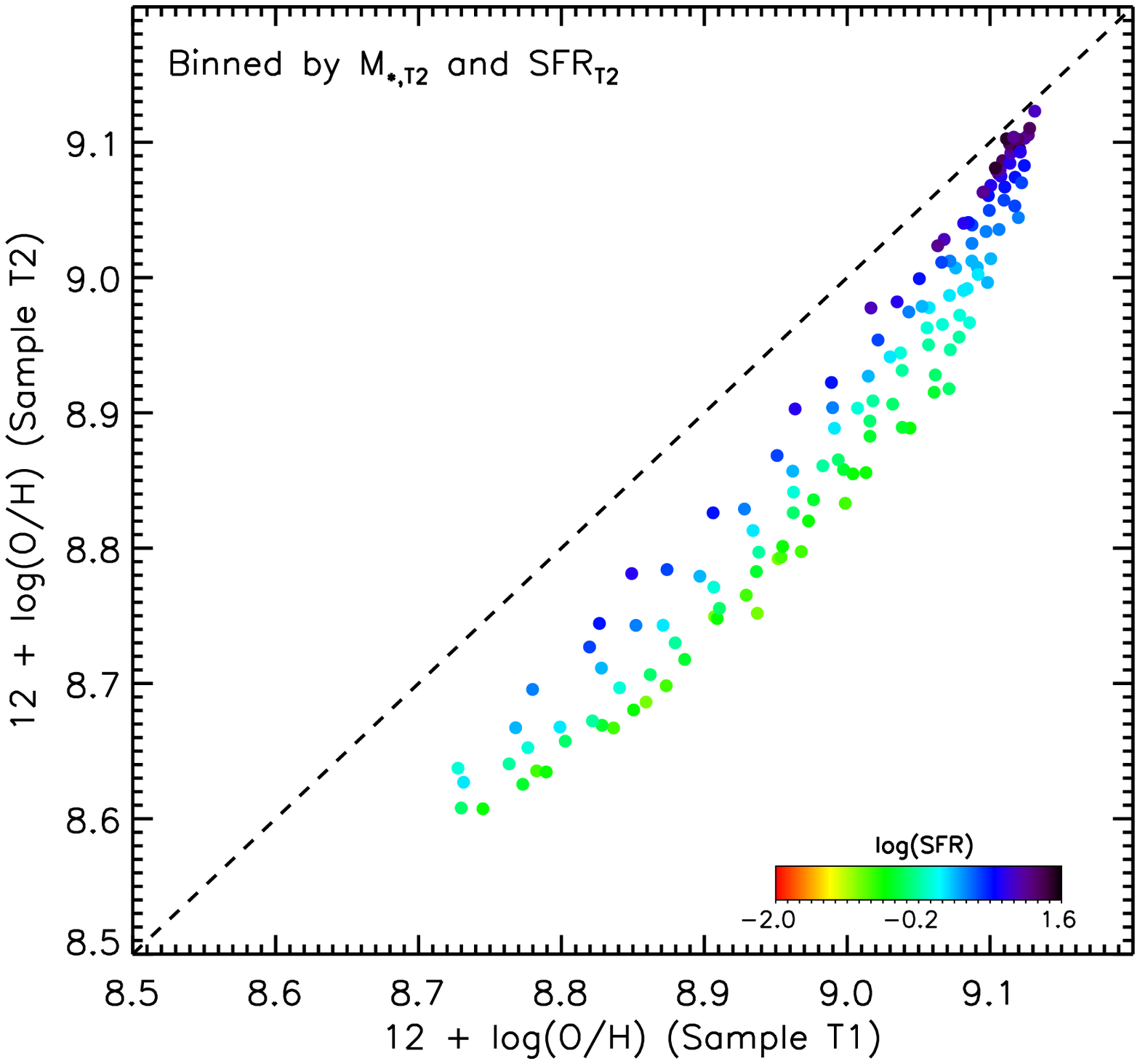}
\caption{A comparison of the oxygen abundances obtained from the two metallicity diagnostics used in this work. Data is binned by $M_{*}$ and the estimation of SFR described in Section \ref{Sample T2}. Points are coloured by SFR. The x-axis represents the technique outlined by \citet{M10}, using strong line ratio calibrations from \citet{M08}. The y-axis represents the Bayesian technique outlined by \citet{T04}. There is a clear and systematic over-estimation of $Z$ from the former method relative to the latter. This discrepancy is also more significant for low-SFR galaxies.}
\label{fig:T1T2_abundance_diff}
\end{figure}

The difference in the dependence of metallicity on SFR at high stellar masses in Sample T2 is mainly attributable to the metallicity derivation method chosen. Fig. \ref{fig:T1T2_abundance_diff} shows a comparison of the values of $Z$ obtained from the two methods used. Only the 93,971 galaxies that are present in both samples are included, and these are re-binned by $M_{*}$ and the Sample T2 value of SFR. We can see that for metallicities below $Z\sim 9.1$, the strong line ratio method used for Sample T1 yields systematically higher values of  $Z$ compared to the Bayesian method used for Sample T2. This difference is also larger for galaxies with lower SFR. A similar effect is seen when binning galaxies by the H$\alpha$-based SFRs used for Sample T1.

It is not straightforward to determine which of these two methods is most accurate at estimating oxygen abundance. The calculation of indirect gas-phase metallicities is fraught with complications, and a full discussion on the merits of different methods is well beyond the scope of this work (however, see Appendix A for a brief discussion). Noting that the properties of Sample T1 are already well described by \citet{M10} using their data, we choose to drop further analysis of Sample T1 and focus on Sample T2 in the rest of this work.

\begin{figure*}
\centering
\begin{tabular}{@{}c@{} @{}c@{} @{}c@{}}
\includegraphics[totalheight=0.23\textheight, width=0.33\textwidth]{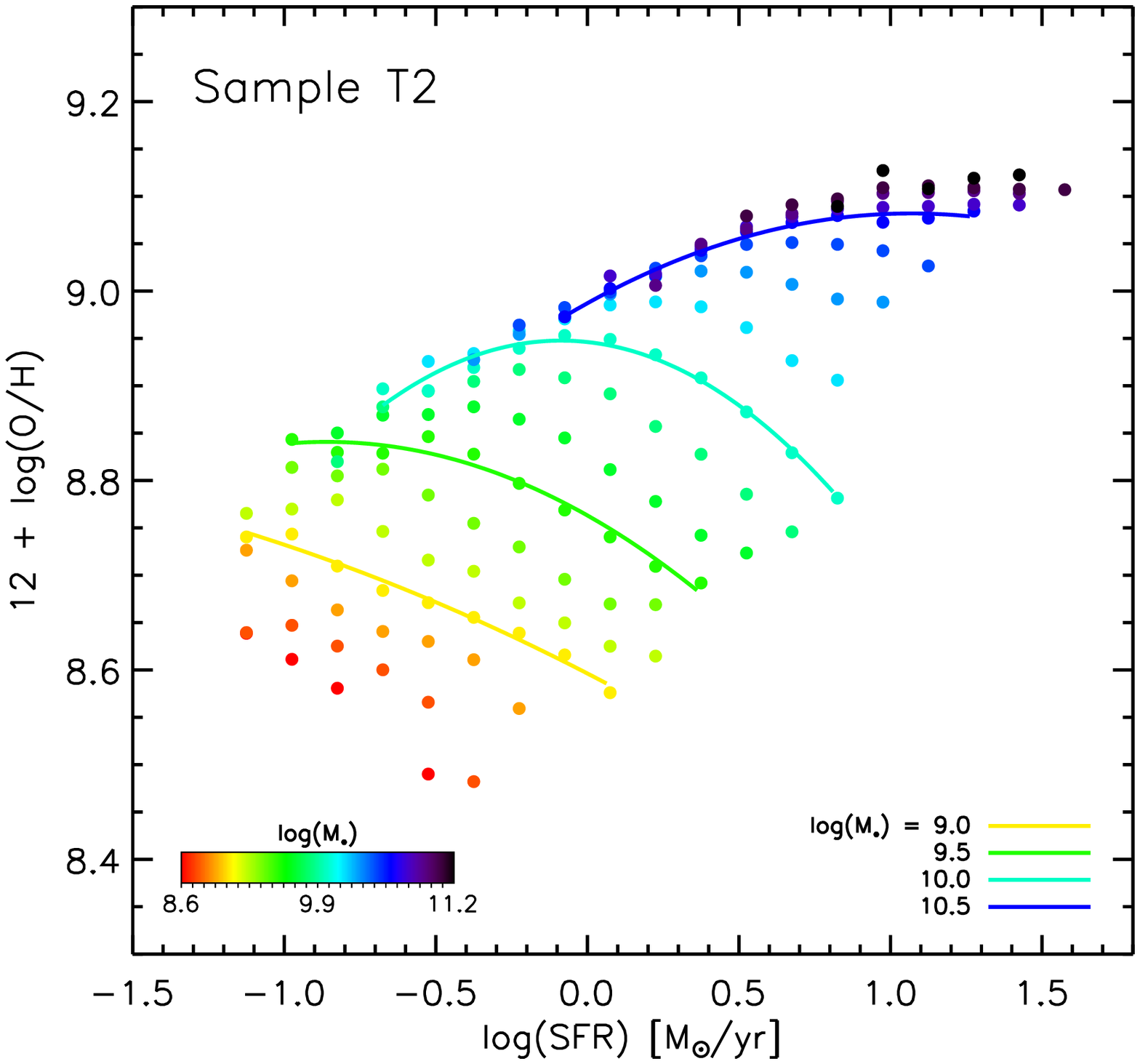} & %, trim=2mm 2mm 2mm 2mm, clip
\includegraphics[totalheight=0.23\textheight, width=0.33\textwidth]{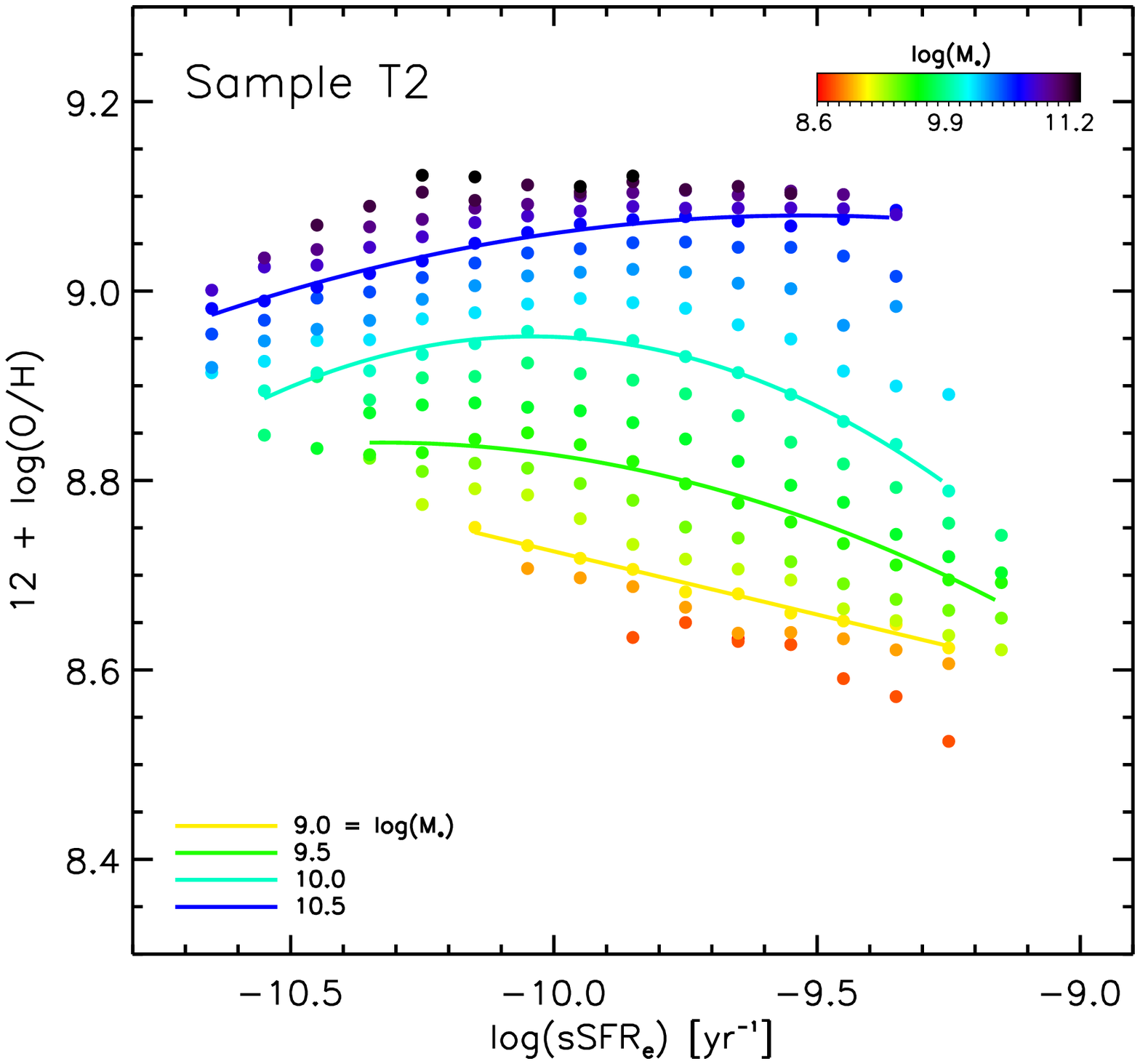} &
\includegraphics[totalheight=0.23\textheight, width=0.33\textwidth]{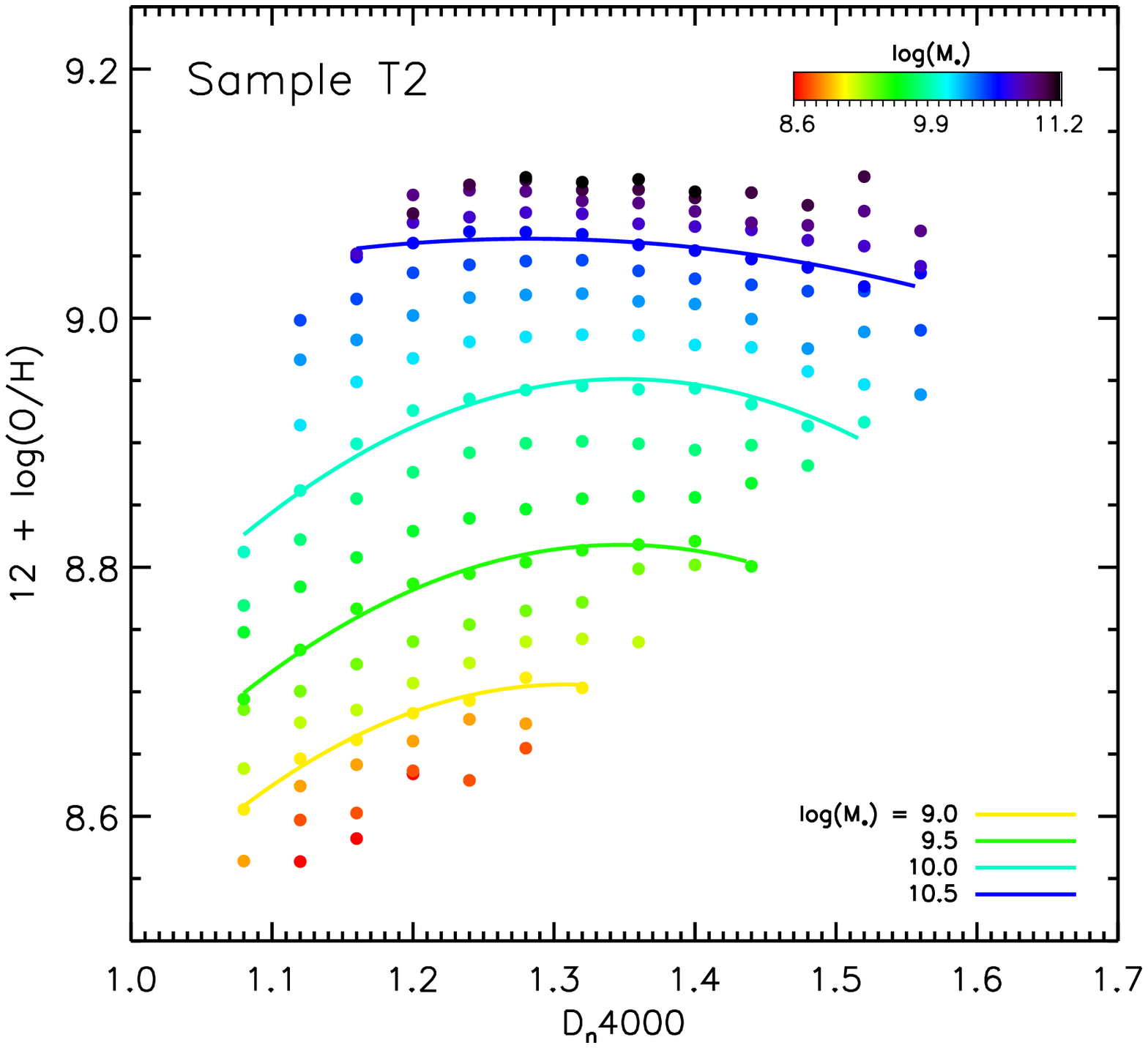} \\
\end{tabular}
\caption{The SFR-$Z$ relation (left panel), (SFR/$M_{*}$)-$Z$ relation (middle panel) and D$_{n}$4000-$Z$ relation (right panel) for Sample T2. Fits to the data for four fixed masses are shown (solid lines). The drop in $Z$ with increasing star formation rate (decreasing D$_{n}$4000) is seen in all three relations. A drop in $Z$ with decreasing star formation rate (increasing D$_{n}$4000) is also seen for high-mass galaxies, representing the same effect seen at high masses in Fig. \ref{fig:M+T_M-Z_mk3}.}
\label{fig:M+M+T_SSFR_Dn4000_mk2}
\end{figure*}

\subsection{The SFR-$Z$ and sSFR-$Z$ relations, as a function of $M_{*}$} \label{sec:The SFR-Z relation}
The SFR-$Z$ relation for Sample T2 is shown in the left panel of Fig. \ref{fig:M+M+T_SSFR_Dn4000_mk2}. Points are coloured by stellar mass, and fits to the relation at four fixed stellar masses are also shown (solid lines).

We can see the drop in $Z$ with increasing SFR for low-mass galaxies, however Sample T2 also exhibits a downturn in metallicity with \textit{decreasing} SFR at high stellar masses. These plots show the same SFR-dependences seen in the $M_{*}$-$Z$ relation, but from another projection onto the $M_{*}$-SFR-$Z$ space. We note that \citet{LL10} carried out an independent investigation of the relation between $M_{*}$, SFR and $Z$ using the SDSS-DR7 online catalogue. They also noticed a significant downturn in metallicity with decreasing SFR in their sample, although they did not study this effect as a function of stellar mass.

Finally, we note that it is actually more sensible to study the dependence of the $M_*$-$Z$ relation on the \textit{specific} star formation rate (SFR/$M_*$, hereafter sSFR$_{e}$), rather than the star formation rate. A dwarf galaxy with a star formation rate of 1 $\textnormal{M}_{\odot} \textnormal{yr}^{-1}$ is a much more `active' system than a giant elliptical galaxy forming stars at the same rate. The specific star formation rate, on the other hand, is a \textit{normalised} quantity, and provides a measure of the present-to-past-averaged star formation rate of the galaxy. Another related quantity is the 4000-\AA ngstrom break strength. This is characterised by the $\textnormal{D}_{n}4000$ index, the average flux from two narrow bands on either side of the break (3850-3950\AA$\!$ and 4000-4100\AA). sSFR estimates based on H$\alpha$ flux are a good measure of the instantaneous star formation rate in a galaxy, whereas $\textnormal{D}_{n}4000$ is more sensitive to stars that have formed over timescales of a few hundred million years to a gigayear. 

The sSFR$_{e}$-$Z$ and $\textnormal{D}_{n}4000$-$Z$  relations for Sample T2 are presented in the middle and right panels of Fig. \ref{fig:M+M+T_SSFR_Dn4000_mk2}. The data has been binned by $M_{*}$ and the sSFR indicator in question. The reader should note that small values of $\textnormal{D}_{n}4000$ correspond to high values of sSFR. 

\begin{figure}
\centering
\begin{tabular}{c}
\includegraphics[totalheight=0.2\textheight, width=0.32\textwidth]{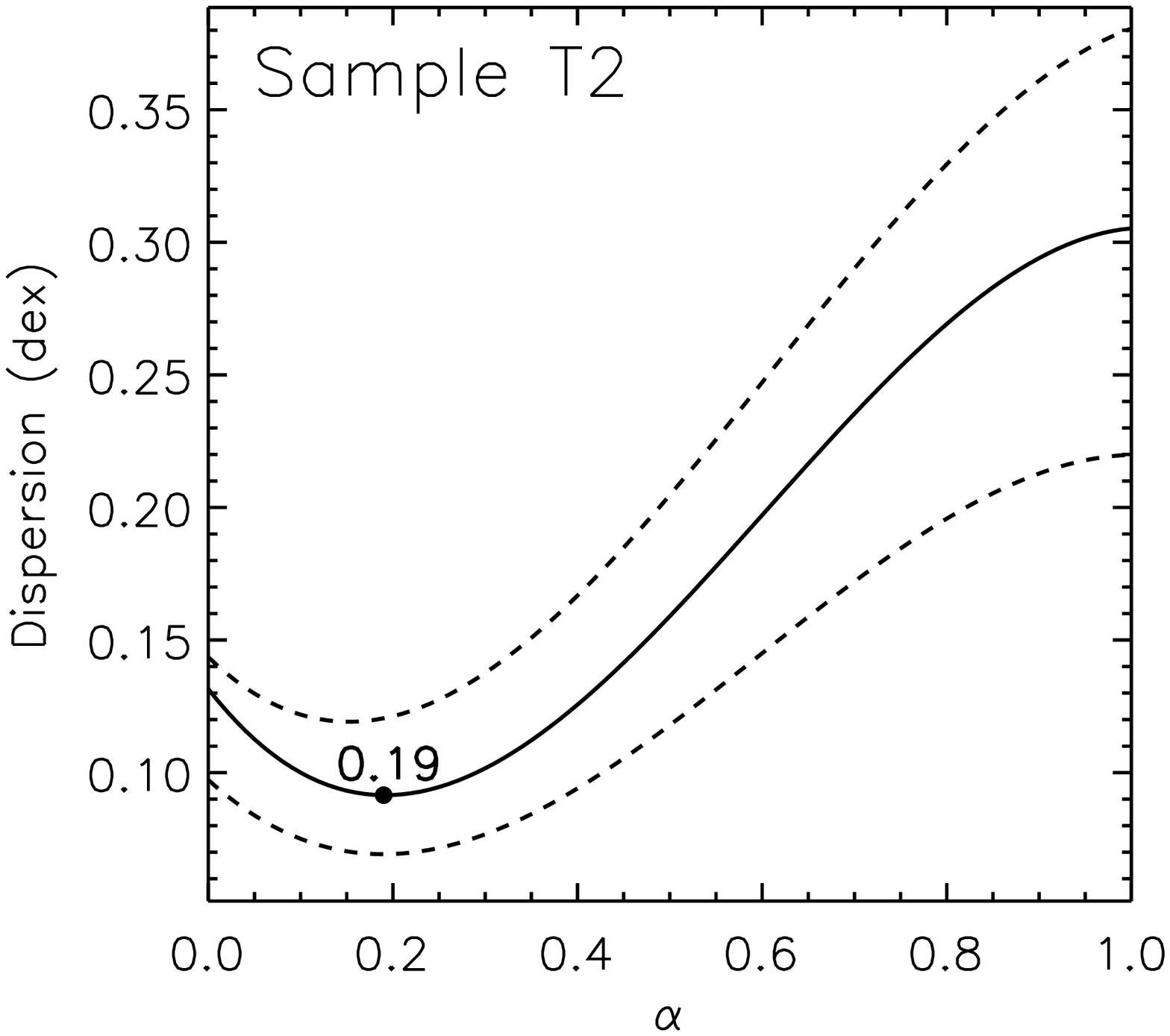} \\
\includegraphics[totalheight=0.2\textheight, width=0.32\textwidth]{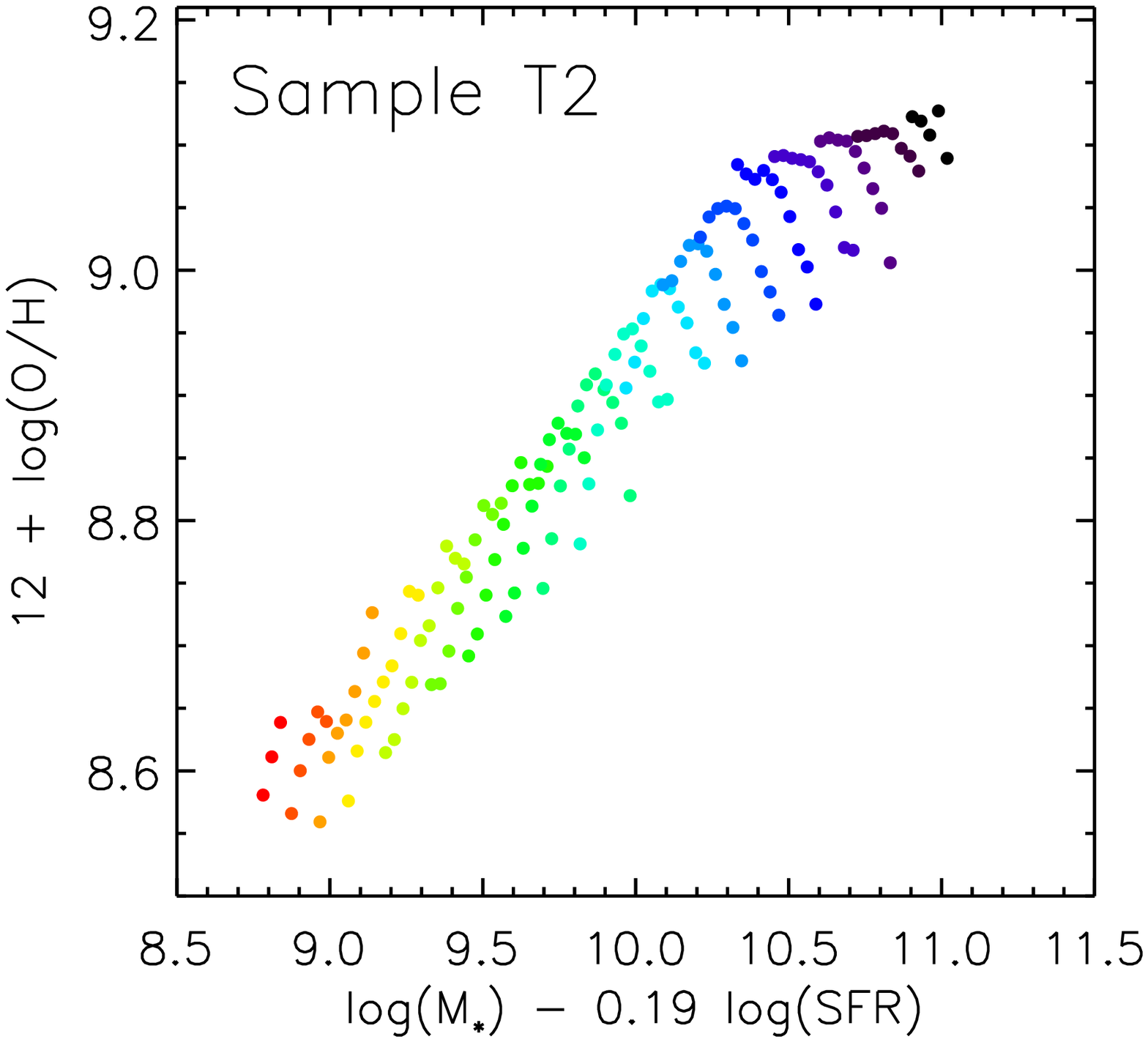} \\
\end{tabular}
\caption{\textbf{Top panel:} The mean dispersion (solid line) in $Z$ for different projections onto the $M_{*}$-SFR-$Z$ space for Sample T2. The 1$\sigma$ scatter in the mean dispersion is also shown (dotted lines). Projections onto the plane are defined by $\alpha$, given in Eqn. \ref{eqn:mu_alpha}. The projection of least scatter for Sample T2 is $\alpha=0.19$ (for Sample T1, this is $\alpha=0.26$). \textbf{Bottom panel:} The projection corresponding to the optimum value of $\alpha$. Only a slight improvement in the dispersion is obtained from this projection compared to the $M_{*}$-$Z$ relation. Points are coloured by SFR, as in Fig. \ref{fig:M+T_M-Z_mk3}.}
\label{fig:Disps}
\end{figure}

It is encouraging that both these relations exhibit similar trends. Metallicity decreases as a function of sSFR for low-mass galaxies and increases as a function of sSFR for high-mass galaxies. When noting that $\textnormal{D}_{n}4000$ (an absorption feature) and sSFR$_{e}$ (computed from emission line fluxes) are independent quantities, the fact that the same trends with metallicity are seen for both indicates that our results are likely to be robust.

\subsection{Projection of least scatter} \label{sec:Projection of least scatter}
If both $M_{*}$ and SFR are correlated with metallicity, then a linear combination of the two may provide a tighter relation with metallicity than the traditional $M_{*}$-$Z$ relation. This was explored by \citet{M10}, who calculated the scatter in median metallicity around their FMR for a series of projections onto the $M_{*}$-SFR-$Z$ space, fixing $Z$ as a principle axis. We modify this method slightly to find the mean dispersion in $Z$ of our binned data for the same 2\textsc{d} projections. Following \citet{M10}, the linear combination of $M_{*}$ and SFR used is denoted by $\mu_{\alpha}$, where

\begin{equation} \label{eqn:mu_alpha}
\mu_{\alpha}=\textnormal{log(}M_{*}\textnormal{)}-\alpha\:\textnormal{log(SFR)}\;\;.
\end{equation}
The free parameter $\alpha$ defines the projection, and can be varied to shift from the $M_{*}$-$Z$ relation ($\alpha=0$) to sSFR$^{-1}$-$Z$ ($\alpha=1$). The corresponding dispersion function for the binned data of Sample T2 can be seen in the top panel of Fig. \ref{fig:Disps}. To obtain this function, the spread in $Z$ was calculated in 0.1 dex bins in $\mu_{\alpha}$, and the mean spread for each projection found. These mean dispersions from $\alpha=0$ to 1 were then fit by a third order polynomial to provide the solid line shown. The 1$\sigma$ spread in the dispersions calculated for each projection was also found and fit in the same way (dotted lines).

It is clear that the $M_{*}$-$Z$ relation ($\alpha=0$) is not the optimum projection for Sample T2. However, the decrease in scatter obtained when using the optimum projection ($\alpha=0.19$) is only slight ($\sim 0.04$ dex). This can be seen by comparing the spread in the $M_{*}$-$Z$ relation in the right panel of Fig. \ref{fig:M+T_M-Z_mk3} with that in the bottom panel of Fig. \ref{fig:Disps}. It is also interesting to note that when using fibre-SFRs, the projection of least scatter for Sample T2 drops to $\alpha=0.03$, very close to the $M_{*}$-$Z$ relation. This is because the spread in star formation rates at low-$M_{*}$ is reduced due to the under-estimation of SFR for nearby galaxies.

The improvement obtained for our Sample T1 is more significant. A projection of $\alpha=0.26$ reduces the dispersion in $Z$ by $\sim0.09$ dex compared to the $M_{*}$-$Z$ relation. The reason for the lack of significant improvement seen for Sample T2 is the inverse dependence of $Z$ on SFR seen at low and high masses. The `u-shape' that the relation therefore forms in the 3\textsc{d} space makes it difficult to find a projection which reduces the overall scatter as much as in Sample T1.

We therefore conclude that, although some improvement to the scatter can be obtained by combining stellar mass and star formation rate in this way, the dispersion around the $M_{*}$-$Z$ relation for Sample T2 is already relatively tight, and that the $M_{*}$-$Z$ relation is suitable for most purposes.

\section{Model Sample} \label{sec:The FMR in L-Galaxies}
In this section, we investigate the relationship between $M_{*}$, SFR and $Z$ in a semi-analytic model of galaxy formation (SAM). This allows us to analyse the detailed evolution of individual galaxies as well as global relations, and to compare the models to observations. Analytic descriptions of physical processes can be self-consistently incorporated into semi-analytic models and then easily adapted, making them more flexible than current SPH simulations. The models also provide large samples of galaxies with predicted stellar masses, metallicities and star formation rates, enabling detailed statistical comparisons with the observations to be made.

\begin{figure}
\centering
\includegraphics[totalheight=0.3\textheight, width=0.46\textwidth]{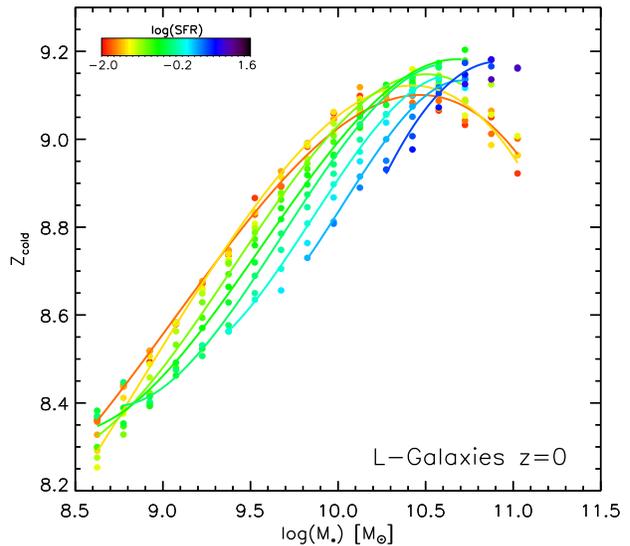}
\caption{The $M_{*}$-$Z$ relation for type 0 and type 1 model galaxies at $z=0$. Binned data (filled circles) are coloured by their SFR, as are the fixed-SFR fits at log(SFR) = -1.725, -1.425, -0.975, -0.675, -0.375, -0.075, 0.225, 0.525. The distinctive feature here is the high-mass turnover for low-SFR galaxies. This turnover causes the reversal of SFR-dependence from low- to high-mass in the model sample, which is also seen in our observational Sample T2.}
\label{fig:L-Gals_ave_z=0}
\end{figure}

The model used in our study is \textsc{L-Galaxies} \citep{G10}, a SAM grafted onto dark matter halo data from the \textsc{Millennium} \citep{S05} 
and \textsc{Millennium-II} \citep{BK09} dark matter N-body simulations. The current model is able to reproduce the observed stellar mass and luminosity functions, as well as the Tully-Fisher and mass-metallicity relations for present-day star-forming galaxies. It also includes updated analytical treatments for gas cooling and stripping, as well as supernova and AGN feedback. The processes important in regulating metallicity that are included in the model are SN-driven outflows with a fixed ejecta speed (this means that metals escape more readily from low-mass galaxies), and gas infall both in the form diffusely accreted pristine gas and gas that was enriched and ejected by the galaxy at an earlier stage. This reintegrated material is returned more quickly into galaxies residing in more massive DM haloes. A detailed description of the analytic recipes used in \textsc{L-Galaxies} can be found in \citet{G10}. The model assumes that the stellar IMF is the same everywhere, and at all epochs.

\begin{figure*}
\centering
\begin{tabular}{@{}c@{} @{}c@{} @{}c@{}}
\includegraphics[totalheight=0.23\textheight, width=0.33\textwidth]{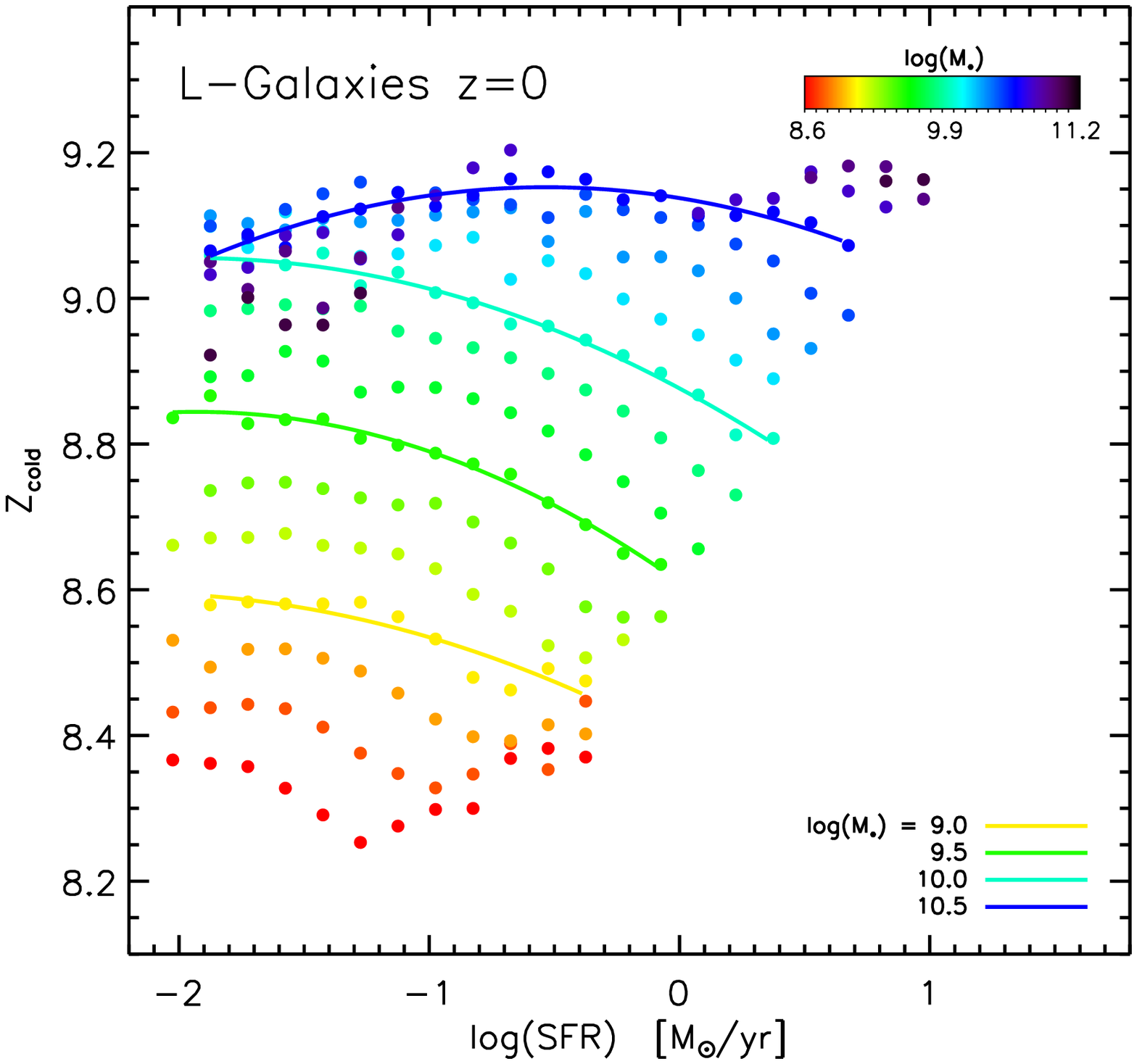} &
\includegraphics[totalheight=0.23\textheight, width=0.33\textwidth]{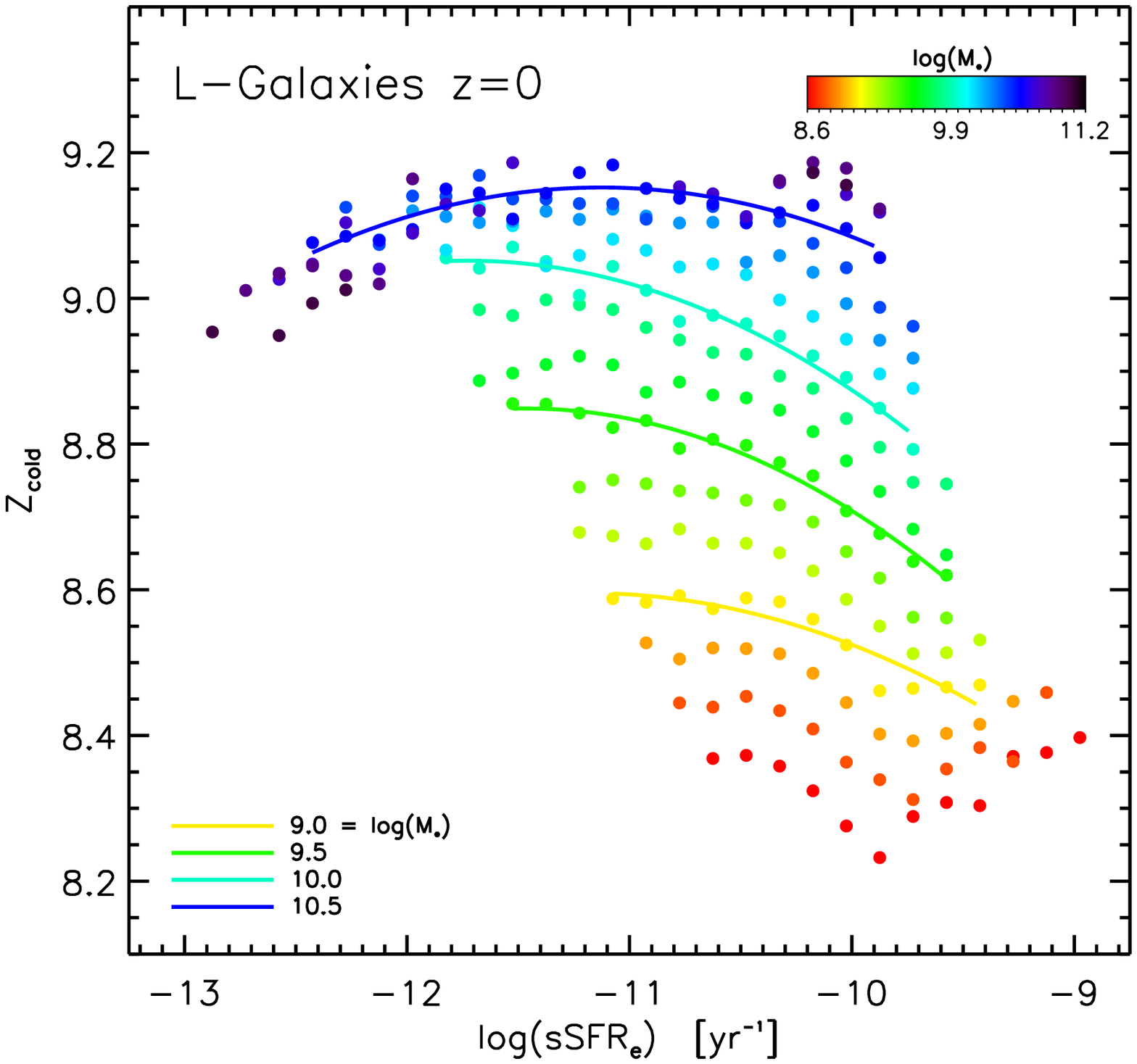} &
\includegraphics[totalheight=0.23\textheight, width=0.33\textwidth]{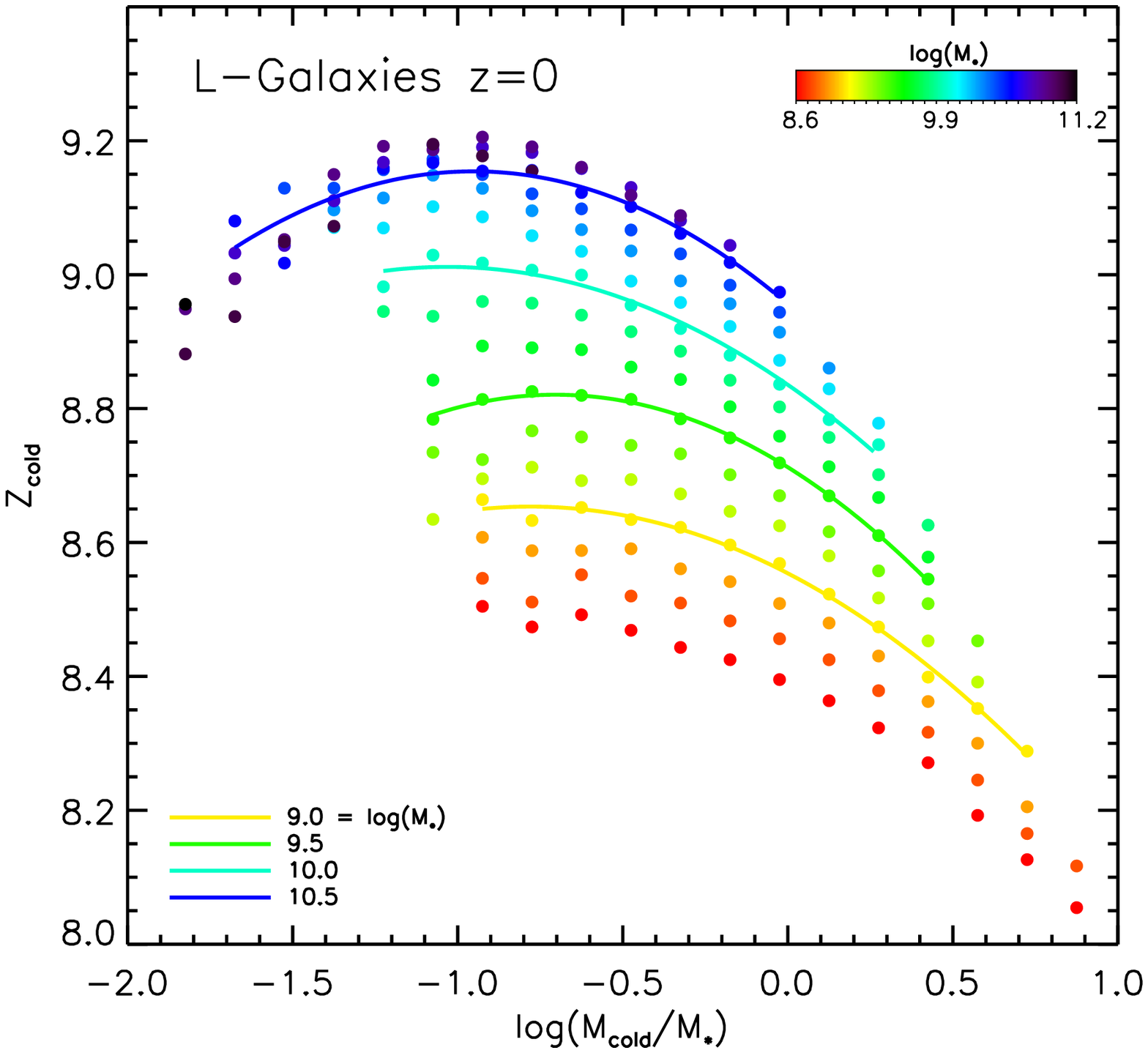} \\
\end{tabular}
\caption{The SFR-$Z$ relation (left panel), (SFR/$M_{*}$)-$Z$ relation (middle panel) and ($M_{\textnormal{cold}}/M_{*}$)-$Z$ relation for the $z=0$ model sample. Fits to the data at four fixed stellar masses are shown (solid lines). The reversal in SFR-dependence from low- to high-masses evident in Fig. \ref{fig:L-Gals_ave_z=0} is again seen. At fixed stellar mass, $Z$ is seen to decrease with increasing SFR (sSFR) at low masses, but decrease with \textit{decreasing} SFR (sSFR) at high masses. Similarly, there is also a clear anti-correlation between gas-to-stellar mass ratio and metallicity at fixed-$M_{*}$ above log($M_{\textnormal{cold}}/M_{*}) \sim -1.0$, but a direct correlation at lower gas-to-stellar mass ratios. These results are complimentary to those seen for our observational Sample T2 in Fig. \ref{fig:M+M+T_SSFR_Dn4000_mk2}.}
\label{fig:L-Gals_SFR-Z_SSFR-Z_z=0}
\end{figure*}

Our $z=0$ model sample comprises 43,767 central galaxies extracted from the Millennium Database provided by the German Astrophysical Virtual Observatory (GAVO)\footnote{available at; \textit{http://www.g-vo.org/Millennium}}. We use the catalogues generated by running the \textsc{L-Galaxies} code on the dark matter halo merger trees from the \textsc{Millennium-II} simulation, which is able to resolve DM halos down to a halo resolution limit of $1.89\times 10^{8} \textnormal{M}_{\textnormal{\astrosun}}$. Galaxies were selected by stellar mass ($8.6\leq\textnormal{log(}M_{*}\textnormal{)}\leq11.2$) and star formation rate ($-2.0\leq\textnormal{log(SFR)}\leq1.6$) to span the same region of parameter space as our observational samples. A WMAP1 cosmology \citep{S03} with the following parameters is assumed: $\Omega_{\textnormal{m}}=0.25$, $\Omega_{\textnormal{b}}=0.045$, $\Omega_{\Lambda}=0.75$, $n_{s}=1$, $\sigma_{8}=0.9$ and $\textnormal{H}_{0}=73$ km s$^{-1}$ Mpc$^{-1}$.

Both type 0 and type 1 central galaxies are included in the model sample. Type 0 galaxies are those lying at the centres of their dark matter haloes. These galaxies accrete material that cools from the surrounding halo, as well as satellite systems that sink to the centre of the halo as a result of dynamical friction. Type 1 galaxies are satellite systems embedded within a larger halo. In the current implementation of \textsc{L-Galaxies}, these galaxies still accrete cold gas from their surrounding self-bound `subhalo'. Over time, the dark matter and gas in the subhalo is tidally stripped and accretion rates decline. The Millennium Database also contains so-called type 2 satellite galaxies, which have lost their surrounding host DM haloes through tidal stripping. These galaxies generally have no ongoing star formation and are therefore not included in our analysis.

\section{Model Results} \label{Model Results}
\subsection{The relation between $M_{*}$, $\textnormal{SFR}$ and $Z$ at $z=0$} \label{sec:The relation between $M_{*}$, SFR and $Z$ at $z=0$}
We bin the galaxies in our sample  by $M_{*}$ and SFR in the same way as was done for our observations. Only bins containing $\geq 25$ galaxies are included in the model plots  -- half the number required in the observational samples. This is done in order to expand the dynamic range in the plots, but none of our results depends on this choice.

The \textit{total cold gas-phase metallicity} for each bin is calculated. This is given by the ratio of mass in metals to total cold gas mass, normalised to the solar metal abundance\footnote{A change in the value of $Z_{\textnormal{\astrosun}}$ down to that preferred by \citet{A09} ($Z_{\textnormal{\astrosun}}=8.69$) does not effect the slope of the relation or the distribution as a function of SFR (see also, \citealt{Be07}). We therefore choose to stick with the value of $Z_{\textnormal{\astrosun}}=9.0$ used by \citet{G10} for consistency.}: $Z_{\textnormal{cold}}=9.0+\textnormal{log}(M_{\textnormal{cold,Z}}/M_{\textnormal{cold}}/0.02)$.

\begin{figure}
\centering
\includegraphics[totalheight=0.3\textheight, width=0.46\textwidth]{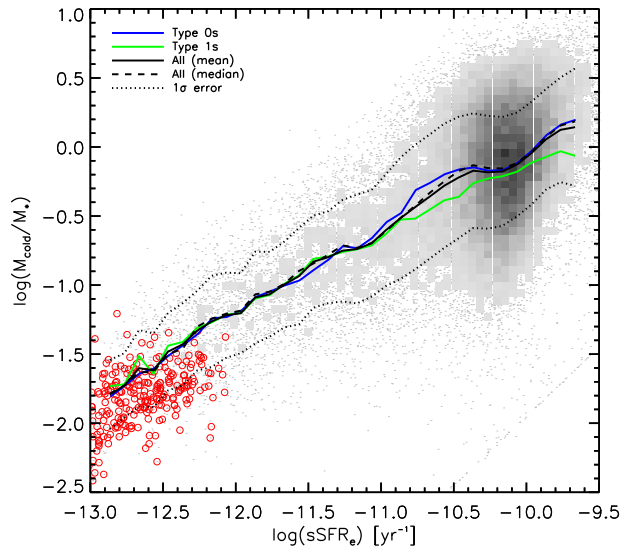} %[totalheight=0.3\textheight, width=0.46\textwidth]
\caption{The sSFR-($M_{\textnormal{cold}}/M_{*}$) relation for our $z=0$ model sample. The relation for type 0 galaxies (blue line) and type 1 galaxies (green line) is shown, as well as the mean (solid black line) and median (dashed black line) relation for the full sample (shown in grey). The 1$\sigma$ spread around the mean is shown as dotted lines. The low-sSFR galaxies that form the high-mass turnover in the model $M_{*}$-$Z$ relation are indicated by red circles. These galaxies tend to have lower-than-average gas-to-stellar mass ratios for their sSFR.}
\label{fig:sSFR-GasFrac}
\end{figure}

The $M_{*}$-$Z$ relation for our $z=0$ model sample is shown in Fig. \ref{fig:L-Gals_ave_z=0}. We see a positive correlation between stellar mass and metallicity, as well as a segregation of this relation by SFR. Interestingly, this segregation is similar to that seen in Sample T2, with a reversal in the SFR-dependence from low to high masses. Another way of stating this result is that there is a `turnover' at high stellar masses in the $M_{*}$-$Z$ relation for low-SFR galaxies. Such a feature is hinted at in Sample T2, but we note that the number of real massive galaxies with emission
lines that are strong enough to measure metallicity and that are not dominated by AGN emission is quite small. This may explain the relative weakness of the feature in the observations compared to the model. The stellar mass at which the turnover occurs is $\sim 10^{10.4} \textnormal{M}_{\textnormal{\astrosun}}$, again in quite good agreement with what is seen in Sample T2. It should be noted that the same turnover is seen when only type 0 galaxies are included in the model sample, and so is not a consequence of environmental effects.

\begin{figure}
\centering
\includegraphics[totalheight=0.34\textheight, width=0.46\textwidth]{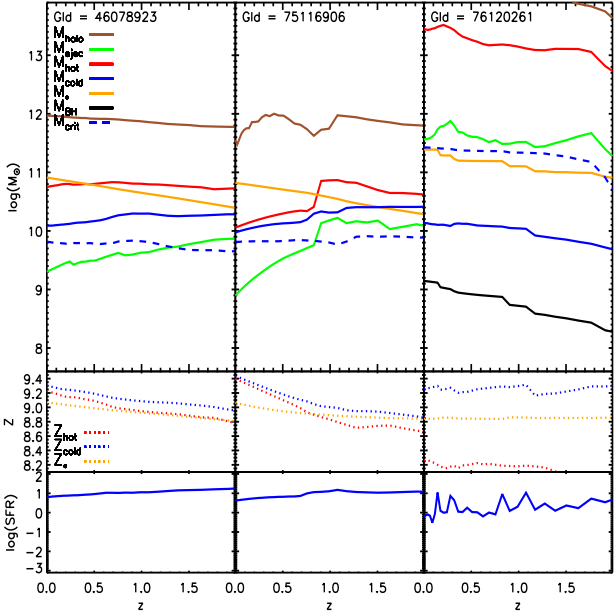}
\caption{The evolution from $z=2$ to 0 in mass (top panels), metallicity (middle panels) and SFR (bottom panels) for three typical galaxies from the high-mass, high-$Z$ sub-sample. Galaxies in this sub-sample tend to undergo gradual enrichment of their cold gas phase over time. The galaxy IDs for these galaxies from the Millennium Database are provided at the top of each panel. }
\label{fig:highZ_specgals}
\end{figure}

The SFR-$Z$ and the sSFR-$Z$ relations are shown in the left and middle panels of Fig. \ref{fig:L-Gals_SFR-Z_SSFR-Z_z=0}. The $\textnormal{D}_{n}4000$ index is not available for our model galaxies, so only sSFR$_{e} = \textnormal{SFR}/M_{*}$ is used. As in our observations, $Z$ depends on star formation at fixed stellar mass. As in Sample T2, the trend for high-mass galaxies is the opposite to that seen at low masses. The metallicities of galaxies of mass $\gtrsim 10^{10.5} \textnormal{M}_{\textnormal{\astrosun}}$ decrease with decreasing SFR (sSFR).

In the right panel of Fig. \ref{fig:L-Gals_SFR-Z_SSFR-Z_z=0} we plot a new relation, that between the gas-to-stellar mass ratio ($M_{\textnormal{cold}}/M_{*}$) and gas-phase metallicity for the same model galaxies. Because cold gas is the fuel for ongoing star formation in a galaxy, the gas-to-stellar mass ratio should correlate with the enrichment of the interstellar medium. Indeed, this is what is seen in our model sample. The average $M_{\textnormal{cold}}/M_{*}$ decreases with stellar mass and with $Z_{\textnormal{cold}}$ for galaxies with stellar masses less than $\sim 10^{10.5}\textnormal{M}_{\textnormal{\astrosun}}$. At higher stellar masses, we again see a turnover towards low metallicities at low $M_{\textnormal{cold}}/M_{*}$. We conclude that the low-sSFR galaxies that contribute to the high-mass turnover in the $M_{*}$-$Z$ relation also tend to have lower $M_{\textnormal{cold}}/M_{*}$ than other galaxies of a similar mass. This is also seen in the sSFR-($M_{\textnormal{cold}}/M_{*}$) relation, shown in Fig. \ref{fig:sSFR-GasFrac}, where these galaxies are indicated by red circles.

Currently, there is only limited data available on the gas fractions of high-mass galaxies with low star formation rates. However, with the ongoing development of the \textsc{Gass} (GALEX Arecibo SDSS Survey, \citealt{C10}) and \textsc{ColdGass} (CO Legacy Database for GASS, \citealt{S11}) programmes, we will soon have a significant number of gas-to-stellar mass ratios for high-mass SDSS galaxies with which to compare. It will be interesting to see if the relationship between $M_{\textnormal{cold}}/M_{*}$, SFR and $Z$ presented in Figs. \ref{fig:L-Gals_SFR-Z_SSFR-Z_z=0} and \ref{fig:sSFR-GasFrac} is also found in these observations.

\subsection{Metallicity evolution in model galaxies} \label{sec:Enrichment evolution}
In this section, we study the origin of the turnover in the model $M_{*}$-$Z$ relation seen in Fig. \ref{fig:L-Gals_ave_z=0}. We do this by splitting the high-mass end of the sample into low metallicity and high metallicity sub-populations and studying differences in their evolutionary histories. We extracted two high-mass ($M_{*} > 10^{10.8}\textnormal{M}_{\textnormal{\astrosun}}$) sub-samples: a high-$Z$ ($Z\geq9.2$) sub-sample containing 134 galaxies, and a low-$Z$ ($Z\leq9.0$) sub-sample containing 136 galaxies. The mass, metallicity and SFR evolution of these two sub-samples was then compared.
% of these galaxies

\begin{figure}
\centering
\includegraphics[totalheight=0.34\textheight, width=0.46\textwidth]{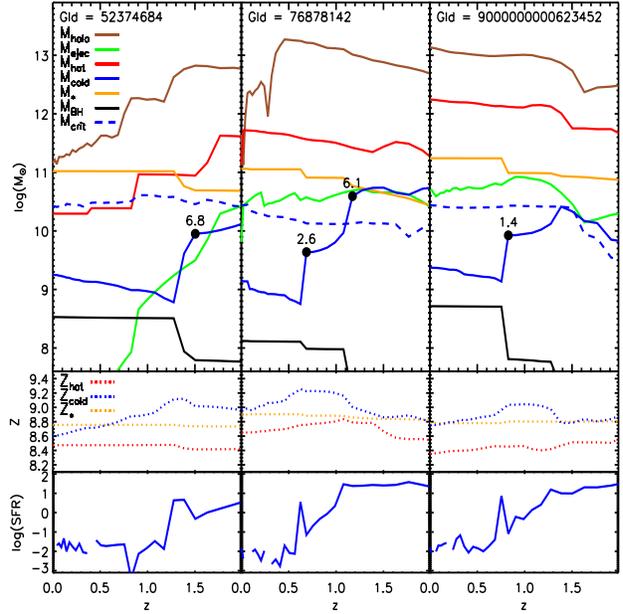}
\caption{The evolution from $z=2$ to 0 in mass (top panels), metallicity (middle panels) and SFR (bottom panels) for three typical galaxies from the high-mass, low-$Z$ sub-sample. These galaxies show the gradual dilution of the cold gas phase that is characteristic of the low-$Z$ sub-sample. The Galaxy IDs for these galaxies from the Millennium Database are provided at the top of each panel.}
\label{fig:lowZ_specgals}
\end{figure}

The \textsc{L-Galaxies} model tracks six distinct mass components of galaxies: stellar mass (in the form of a bulge and disc), black hole mass, 
cold gas mass (ISM), hot gas mass (ICM), ejected gas mass (IGM) and halo stars (producing the intra-cluster light). Mass and metals can pass between these components along pre-defined routes, depending on the processes taking place. The top three panels of Fig. \ref{fig:highZ_specgals} show the time evolution of these mass components for three representative galaxies from the model with high stellar masses and high metallicities. In the middle panels, we show the time evolution of the stellar and gas-phase metallicities of the same galaxies. The bottom panels show the time evolution of their star formation rates. 

The left panels display a type 0 galaxy in which the stellar mass (solid orange line) has been steadily increasing since redshift two. In this galaxy, the mass of cold gas (solid blue line) is always higher than the critical value required for star formation, $M_{\textnormal{crit}}$ (dashed blue line).\footnote{Note that the Toomre Q disc instability criterion for star formation \citep{T64} is expressed in \textsc{L-Galaxies} in terms of a critical cold gas \textit{mass} (\citealt{G10}, Section 3.4)} There is a steady, gradual increase in the metallicity of the stellar, cold gas and hot gas components. This is because stars are formed continuously, synthesising and distributing metals throughout the galaxy at a higher rate than the dilution due to the accretion of metal-poor gas. Around 64 per cent of the galaxies in our high-$Z$ sub-sample have formation histories similar to this.

The middle panels show a galaxy that first evolves in a similar way to the galaxy shown in the left panels. It is then accreted onto a more massive DM halo at $z\sim 1.0$, becoming a type 1 object, at which point gas and dark matter begin to be tidally stripped. Some of these type 1 galaxies, like the example shown in the middle panel, then exhibit a sharp increase in gas-phase metallicity. This is because gas accretion onto the galaxy is reduced, but star formation continues, and as a result, metals continue to be dispersed into ever decreasing volumes of hot and cold gas. In other type 1 galaxies in the high-$Z$ sub-sample, the cold gas mass drops below $M_{\textnormal{crit}}$ after being accreted, causing the cold gas metallicity to remain constant thereafter, due to a shut-down in both star formation and galactic gas accretion. Type 1 galaxies with star formation histories of these forms constitute $\sim 30$ per cent of the high-$Z$ sub-sample. 

The right panels of Fig. \ref{fig:highZ_specgals} show a class of central galaxy that formed a bulge and a supermassive black hole at redshifts greater than two. The black hole (solid black line) then grew steadily through so-called `radio mode' accretion of hot gas from the surrounding halo. Such galaxies are not representative of the high-$Z$ sub-sample as a whole (they comprise only  6 per cent of the sub-sample) and are almost certainly classified as AGN and so missing from our observational samples. Nevertheless, we think they are rather interesting. The galaxy makes it into our high-$Z$ model sub-sample due to the fact that it accretes many, many satellites over time (4,141 since redshift two, compared to only 21 and 71 for 
the galaxies in the left and middle panels, respectively). The satellites bring in fresh gas, leading to the very `bursty' star formation history seen in the bottom-right panel of Fig. \ref{fig:highZ_specgals}. We note that this galaxy resides in the center of the 9th most massive DM halo in the whole \textsc{Millennium-II} simulation at $z=0$, and so such a high rate of satellite accretion is perhaps unsurprising.

In conclusion, for the majority of galaxies in the high-$Z$ sub-sample, the dominant process driving metallicity evolution is clearly a \textit{gradual enrichment} of the gas phase due to continuous star formation. At these high masses, outflows are inefficient at removing cold gas and metals from the galaxy. At these high SFRs, galactic infall rates are too low to dilute the ISM. Consequently, these galaxies become increasingly metal-rich with time.

\begin{figure}
\centering
\includegraphics[totalheight=0.3\textheight, width=0.46\textwidth]{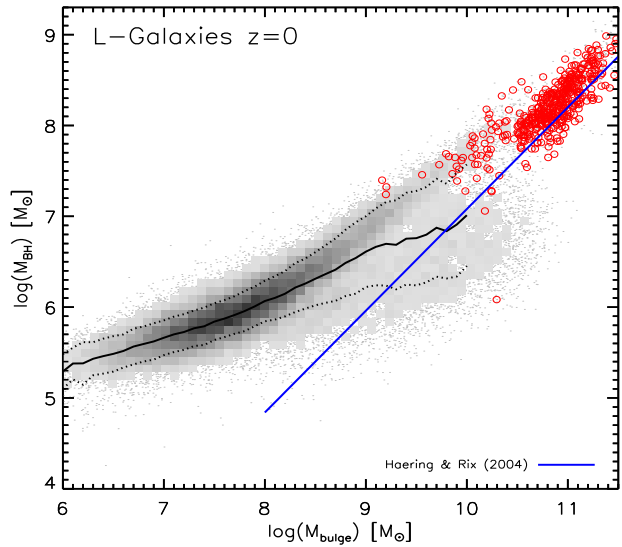} %[totalheight=0.25\textheight, width=0.36\textwidth]
\caption{The $M_{\textnormal{bulge}}$-$M_{\textnormal{BH}}$ relation for our $z=0$ model sample. The full sample is shown in grey, with the mean relation plotted as a solid black line. The 1$\sigma$ spread around the mean is shown as dotted lines. Galaxies in the low-$Z$ sub-sample are highlighted by red circles. These galaxies clearly have higher black hole masses that expected from an extrapolation of the model relation, but lie nicely along the observational relation derived by \citet{HR04}.}
\label{fig:Mbulge-MBH}
\end{figure}

Fig. \ref{fig:lowZ_specgals} paints a rather different picture for the evolution of galaxies in the low-$Z$ sub-sample. All three galaxies have 
undergone dramatic drops in their cold gas masses, coinciding with a merger event at some stage during the past ten gigayears. Mergers with mass ratio less than 10:1 are marked by black dots on the $M_{\textnormal{cold}}$ evolution line. Note that no such mergers occur for the three high-$Z$ examples
in Fig. \ref{fig:highZ_specgals} -- only 7.5 per cent of galaxies in the high-$Z$ sub-sample have undergone a significant merger over the last ten gigayears. A merger with mass ratio 3:1 or less is considered a major merger in the model, causing the destruction of the stellar and gas discs and the transfer of this material to the bulge of the descendant. We see from Fig. \ref{fig:lowZ_specgals} that not only major mergers cause the sudden drop in $M_{\textnormal{cold}}$. Gas-rich minor mergers are also effective at inducing starbursts and the rapid growth of the central SMBH through `quasar mode' accretion. During such events, a black hole can grow by swallowing both cold gas and the smaller black hole of its companion. 92 per cent of the galaxies in the low-$Z$ sub-sample have present-day black holes with masses greater than $10^{8.0} \textnormal{M}_{\textnormal{\astrosun}}$, that were formed through this process. The remaining 8 per cent have either grown their black holes gradually through radio-mode accretion, or do not contain a central SMBH. 

%Although the presence of a large black hole \textit{does not} cause the lower metallicities seen in these galaxies at $z=0$, it is a feature which could help us identify similar galaxies in observations (see Section \ref{sec:Black hole masses}).

\begin{figure*}
\centering
\begin{tabular}{cc}
\includegraphics[totalheight=0.3\textheight, width=0.46\textwidth]{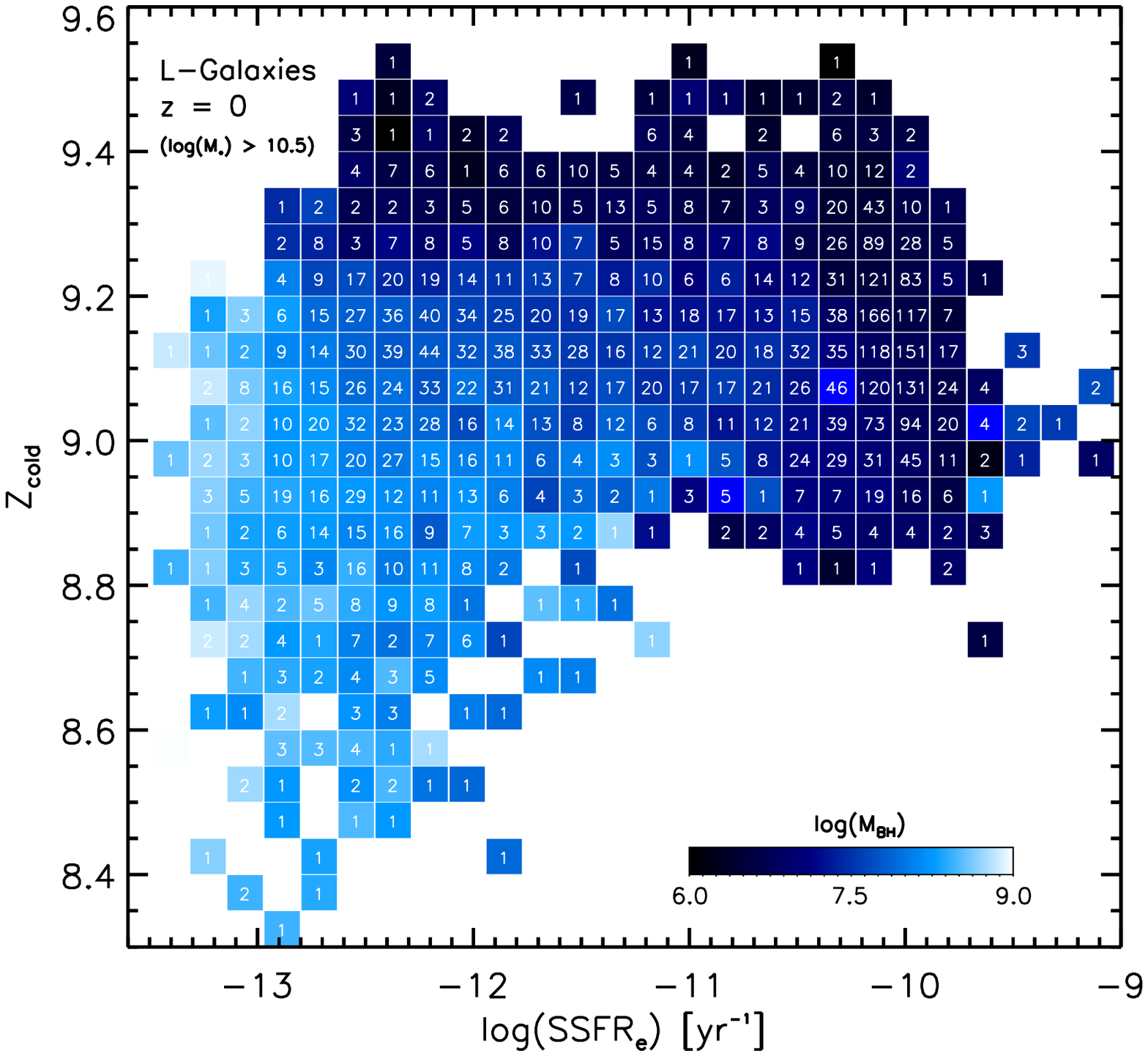} &
\includegraphics[totalheight=0.3\textheight, width=0.46\textwidth]{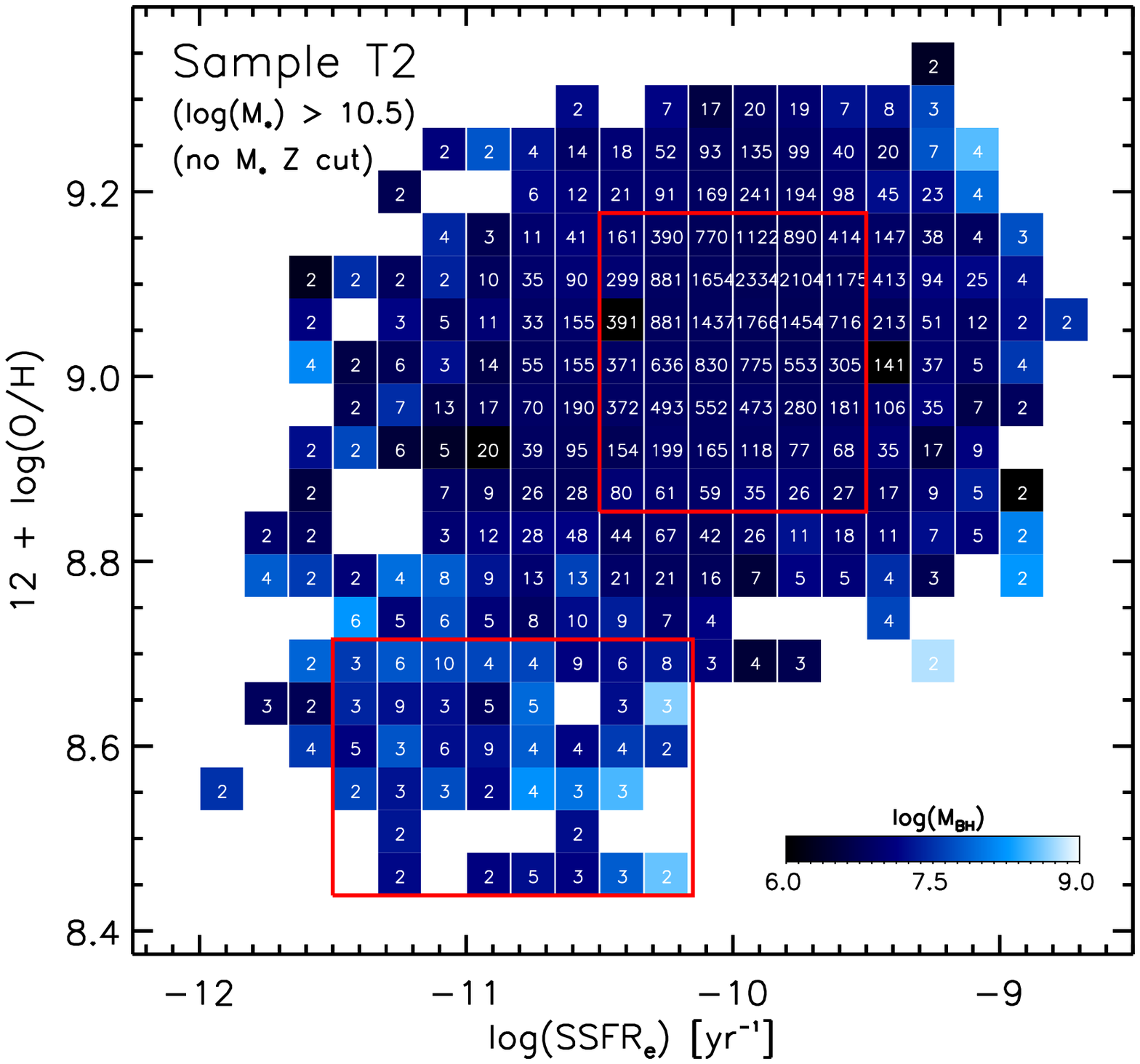}
\end{tabular}
\caption{The (SFR/$M_{*}$)-$Z$ relation for our model sample (left panel) and observational Sample T2 (right panel) for galaxies of stellar mass $> 10^{10.5} \textnormal{M}_{\textnormal{\astrosun}}$. Galaxies had been binned by sSFR$_{e}$ and $Z_{\textnormal{cold}}$, and bins are coloured by the mean black hole mass of the constituent galaxies. The number of galaxies in each bin is also given by white text. The low sSFR, low-$Z$ population clearly have larger central SMBHs in the model sample. A similar dichotomy can be seen in Sample T2, though the distinction is not as clearly defined. The red boxes in the right panel mark the low-$Z$ and high-$Z$ regions shown in Fig. \ref{fig:BHDists}.}
\label{fig:sSFR-Z_BH}
\end{figure*}

We note that although large black holes are a feature of almost all the galaxies in the low-$Z$ sub-sample, they do not \textit{cause} the low metallicities seen in these galaxies at $z=0$. These are instead caused by a cessation in star formation due to the sudden drop in cold gas mass below $M_{\textnormal{crit}}$, followed by accretion of metal-poor gas. This galactic accretion is limited to a low rate by the suppression of cooling from radio mode AGN feedback, allowing it to continue for an extended amount of time without re-igniting star formation. The three galaxies in Fig. \ref{fig:lowZ_specgals} therefore show that \textit{gradual dilution} of the gas phase due to metal-poor infall of gas in the absence of star formation is the main process producing the low-$Z$ sub-sample.

This is an effect that is not seen in hydrodynamic simulations of galaxy evolution such as those carried out by \citet{FD08} and \citet{D11a}. In those models, galaxies quickly fall back into an equilibrium between their infall, outflow and star formation rates after a perturbative event, whereby $\dot{M}_{\textnormal{infall}} = \dot{M}_{*} + \dot{M}_{\textnormal{outflow}}$. Instead, the inclusion of AGN feedback in our model enables galaxies to slowly accrete metal-poor gas for a number of gigayears without forming stars.

We note that it remains to be seen whether the fraction of massive galaxies in the models with very little ongoing star formation and low metallicity reservoirs of cold gas is matched in observations. The low gas fractions of massive galaxies means that the gas is difficult to detect. Obtaining H\textsc{i} maps of a large sample of such objects remains a significant observational challenge.

\begin{figure}
\centering
\includegraphics[totalheight=0.25\textheight, width=0.37\textwidth]{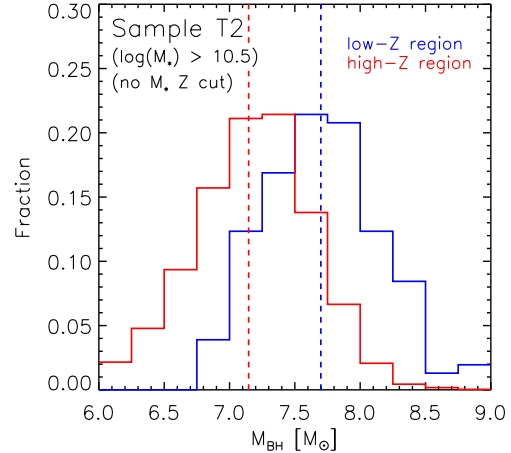} %[totalheight=0.3\textheight, width=0.46\textwidth]
\caption{The distribution of black hole masses in Sample T2 for low-$Z$ (blue histogram) and high-$Z$ (red histogram) galaxies. These are the galaxies marked by the red boxes in Fig. \ref{fig:sSFR-Z_BH}. The mean $M_{\textnormal{BH}}$ of each region is also show by vertical dashed lines. Low-$Z$, low-sSFR galaxies have more massive central black holes in Sample T2, as seen in the model sample. Shifting the boundaries of the two regions somewhat does not affect this result.}
\label{fig:BHDists}
\end{figure}

\begin{figure*}
\centering
\includegraphics[totalheight=0.4\textheight, width=0.58\textwidth]{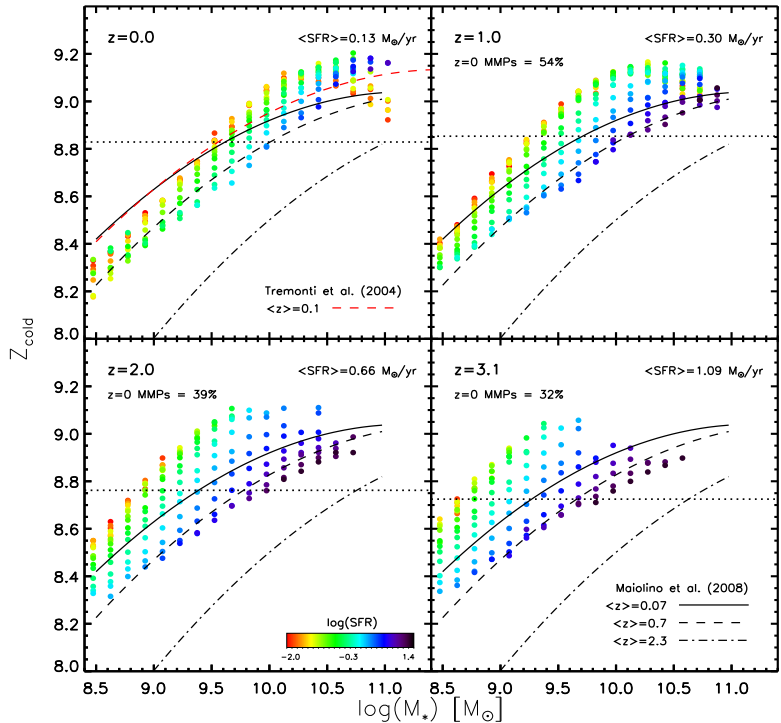}
\caption{The evolution of the $M_{*}$-$Z$ relation in \textsc{L-Galaxies} from $z=3.1$ to $0$. The average metallicity at each redshift is plotted (horizontal dotted lines), and the mean SFR at each redshift is given in the top right of each panel. The percentage of galaxies in each higher redshift sample that have direct descendants in the $z=0$ sample is shown in the top left of each higher redshift panel. Fits to observational data at three different redshifts compiled by \citet{M08} are also shown. These are taken from \citet{KE08} for $<z>=0.07$, \citet{Sa05} for $<z>=0.7$ and \citet{E06c} for $<z>=2.3$. A fit to the \citet{T04} $<z>=0.1$ relation (red dashed line) is also shown in the $z=0$ panel.}
\label{fig:L-Gals_M-Z_allz}
\end{figure*}

\subsection{Bulge and black hole masses} \label{sec:Black hole masses}
As discussed in the previous section, one clear distinguishing feature of massive, low-$Z$ galaxies in the model, aside from a low gas-to-stellar mass ratio, is the presence of a massive bulge and massive central black hole formed during a merging event. This is illustrated by the $M_{\textnormal{bulge}}$-$M_{\textnormal{BH}}$ relation in Fig. \ref{fig:Mbulge-MBH}, where the galaxies in the low-$Z$ sub-sample are highlighted by red circles. Interestingly, it is these galaxies that lie closest to the observational relation found by \citet{HR04}. A significant fraction of intermediate to massive galaxies with smaller central black holes have likely had their bulges grown through secular processes (\citealt{Sh11}; Shankar, in prep.). Although massive central black holes are not the cause of the low metallicities in the low-$Z$ sub-sample, they are an associated feature that we can look for in the currently available observational data.

In the left panel of Fig. \ref{fig:sSFR-Z_BH}, specific star formation rate is plotted against gas phase metallicity for all our high mass model galaxies. Galaxies are binned by sSFR$_{e}$ and $Z_{\textnormal{cold}}$, and the bins are coloured by the mean black hole mass of the galaxies in each bin. The number of galaxies in each bin is also indicated on the plot in white text. It is clear that low-sSFR, low-$Z$ galaxies have the largest central black holes in the model.

The right panel of Fig. \ref{fig:sSFR-Z_BH} shows the same plot for an adapted version of our observational Sample T2 (see Appendix B for details). Black hole masses were estimated via the measured stellar velocity dispersion using the $M_{\textnormal{BH}}$-$\sigma$ relation provided by \citet{T02}:

\begin{equation} \label{eqn:Mbh-veldisp}
\textnormal{log}(M_{\textnormal{BH}}) = \alpha + \beta\textnormal{log}(\sigma/\sigma_{0})\;\;,
\end{equation}
where $\alpha = 8.13$, $\beta = 4.02$, $\sigma_{0} = 200$km/s and $M_{\textnormal{BH}}$ is in units of $\textnormal{M}_{\textnormal{\astrosun}}$. We can see that the observational data also contains a low-$Z$, high-$M_{\textnormal{BH}}$ population, though the distinction between this population and the majority of the sample is less clear than in the model.

The dichotomy in the observations is more clearly seen in Fig. \ref{fig:BHDists}, where the distribution of black hole masses is shown for galaxies contained within the two red boxes marked in Fig. \ref{fig:sSFR-Z_BH}. The black hole population is clearly shifted to higher masses in the low-$Z$ region (blue histogram), compared to the high-$Z$ region (red histogram). 24 per cent of galaxies within the low-$Z$ region have black holes of mass $\geq 10^{8.0} \textnormal{M}_{\textnormal{\astrosun}}$. This is only true for 2.7 per cent of galaxies in the high-$Z$ region. A correlation between low metallicities, low-sSFRs and high black hole masses therefore seems present in both our model and observational sample.

\subsection{Evolution of the $M_{*}$-$Z$ relation out to $z\sim 3$} \label{sec:Evolution of MZR}
In this section, we analyse the evolution of the $M_{*}$-$Z$ relation out to $z\sim 3$. Observations have shown a clear evolution in the $M_{*}$-$Z$ relation \citep{M08} and $M_{*}$-SFR relation \citep{N07a,N07b,K10} with look-back time, and such evolution is required for the Mannucci FMR to remain fixed out to $z\sim 2.5$.

In order to investigate this in \textsc{L-Galaxies}, three supplementary, identically selected samples of type 0 and 1 galaxies were extracted from the database at redshifts $z=1.0$, 2.0 and 3.1. These samples contain 63,745, 63,017 and 50,558 galaxies, respectively, within the parameter space of interest.

The evolution of the $M_{*}$-$Z$ relation from $z=3.1$ to 0 in the model is plotted in Fig. \ref{fig:L-Gals_M-Z_allz}. The mean SFR at each redshift is also given in the top right of each panel. The SFR is seen to evolve strongly with redshift, dropping by $\sim 0.9$ dex from $z=3.1$ to 0.0 at fixed mass, in line with the drop observed by \citet{N07a,N07b} and \citet{K10}. However, there appears to be nearly no evolution in metallicity at fixed stellar mass at all in the model, contrary to observations. The present day $M_*$-$Z$ relation agrees well with those of \citet{T04} and \citet{KE08} at $z\sim 0$, but the discrepancy with observations becomes increasingly pronounced towards higher redshifts. This suggests that chemical enrichment in the model galaxies proceeded too rapidly at early times.

\begin{figure}
\centering
\includegraphics[totalheight=0.3\textheight, width=0.46\textwidth]{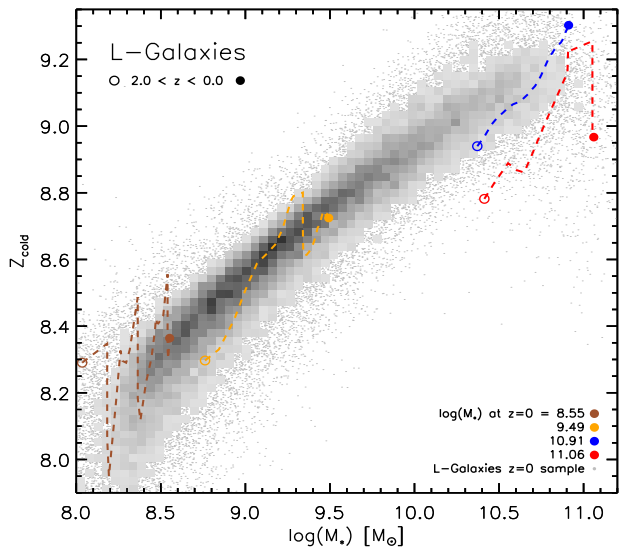}
\caption{The evolution of four model galaxies from our $z=0$ sample from redshift 2 to 0. Galaxies of final stellar mass log$(M_{*}) = 8.55$, 9.49, 10.91 and 11.06 are shown in brown, orange, blue and red, respectively. The two higher mass examples are \textsc{GId}46078923 (blue) and \textsc{GId}76878142 (red) from the high-mass sub-samples shown in Figs. \ref{fig:highZ_specgals} and \ref{fig:lowZ_specgals}. The full present day population is also shown in grey for comparison. We can see that galaxies tend to evolve along the $z=0$ relation at all masses, but that the nature of this evolution is different for different mass intervals.}
\label{fig:L-Gals_IndGalsEvo}
\end{figure}

In order to diagnose whether this hypothesis is correct, we plot the enrichment history from $z=2$ to the present day of four representative galaxies from the $z=0$ model sample in Fig. \ref{fig:L-Gals_IndGalsEvo}. The $M_{*}$-$Z$ relation for the entire $z=0$ sample is also plotted in grey. This is a reasonable test, as a large number of galaxies in the $z=0$ sample have main progenitors present in the higher redshift samples (the percentage of galaxies that lie on the most massive progenitor (MMP) branch of a $z=0$ galaxy is indicated in the top left corner of each higher redshift panel in Fig. \ref{fig:L-Gals_M-Z_allz}).

We can again see in Fig. \ref{fig:L-Gals_IndGalsEvo} the two types of behaviour at high stellar masses described in Section \ref{sec:Enrichment evolution}, with both $Z_{\textnormal{cold}}$ and stellar mass gradually increasing with time for the blue (high-SFR) galaxy, and metallicity decreasing with time without any associated increase in stellar mass for the red (low-SFR) galaxy. At intermediate masses, stellar mass and metallicity increase together, punctuated by periods when $Z_{\textnormal{cold}}$ drops suddenly, due to an episode of enhanced gas accretion, such as a gas rich merger with a metal-poor satellite.

At low masses, the metallicity evolution appears much more erratic. Drastic fluctuations in $Z_{\textnormal{cold}}$ may occur when outflows become extreme enough to drive most of the gas out of the galaxy. This is the case for the lowest mass galaxies in the model, as a result of the SN feedback scheme that has been implemented in \textsc{L-Galaxies}.

We can see from the grey points in Fig. \ref{fig:L-Gals_IndGalsEvo} that the scatter in the $M_{*}$-$Z$ relation at $z=0$ increases towards lower masses as a consequence of this stochastic mode of evolution. We note that it is unclear whether this feature is seen observationally. \citet{L06} have used a sample of 27 nearby dwarf galaxies to argue that the scatter in the observed local $M_{*}$-$Z$ relation remains roughly constant down to stellar masses of $\sim 10^{6.5} \textnormal{M}_{\textnormal{\astrosun}}$. If confirmed with larger samples, this might suggest that, unlike metals, cold gas is not driven very effectively out of low mass galaxies by supernovae \citep{MLF99}.

The main point to take away from  Fig. \ref{fig:L-Gals_IndGalsEvo} is that galaxies evolve along the present-day $M_{*}$-$Z$ relation in the model, rather than from much lower metallicities, as suggested by observations. This suggests that it is indeed an overly rapid enrichment of the cold gas phase of galaxies before redshift three that is causing the lack of evolution thereafter in the model.

One possible solution to the problem is that accreted gas is `hung-up' in the atomic phase of the ISM for some time, before it is able to reach high enough densities to form the molecular clouds in which stars are formed \citep{GTK09,GK10}. Recently, \citet{F10} implemented simplified prescriptions for the formation of molecular gas in galaxies in the \textsc{L-Galaxies} code, to study scaling relations between gas and stars in local galaxies. The recipes have, however, not yet been fully applied to models running on the higher resolution \textsc{Millennium-II} simulation, or to the more recent \citet{G10} version of \textsc{L-Galaxies}, which contains stronger and more mass sensitive feedback. This will be the focus of future work.

\section{Discussion} \label{sec:Discussion}
In Section \ref{sec:The FMR in observations}, we have shown that our preferred observational sample, Sample T2, exhibits a reversal in the dependence of metallicity on star formation rate from low to high stellar mass. At low masses, low-SFR galaxies have higher metallicities compared to more star forming galaxies. At high masses, they have lower metallicities. This observation alone could be explained by considering mass-dependent, metal-rich outflows. At lower masses ($\sim 10^{9} \textnormal{M}_{\textnormal{\astrosun}}$), where outflows are more effective, low-SFR galaxies produce relatively fewer SNe, disrupting the ISM less and blowing away metals less efficiently (e.g. \citealt{MLF99}). At higher masses ($\gtrsim 10^{10.2} \textnormal{M}_{\textnormal{\astrosun}}$), where outflows become increasingly weak, low-SFR galaxies simply produce less metals and so under-enrich the ISM relative to more actively star forming galaxies.

However, our model also shows a decrease in metallicity with stellar mass at fixed-SFR above $\sim 10^{10.4} \textnormal{M}_{\textnormal{\astrosun}}$. Such a feature \textit{cannot} be explained by mass-dependent outflows alone. If the turnover in the $M_{*}$-$Z$ relation is indeed real, then additional physical mechanisms must be at play.

Our model points to metal-poor galactic infall at high-mass as an explanation. Those high-$M_{*}$ galaxies with low gas-phase metallicities are known to have undergone gradual dilution of their gas phases after a merger event which shut-down further star formation. The restriction in the amount of infall by AGN feedback allowed these galaxies to \textit{slowly} dilute their ISM, without accreting enough gas for star formation to resume. Correlations between these galaxies and the high-$M_{*}$, low-$Z$ galaxies in our observational sample (namely, their large black hole masses) imply that such a dilution process could also be involved in shaping the $M_{*}$-$Z$ relation in the real Universe (see also Appendix B).

There are, however, two factors hampering this interpretation. First, the dependence of $Z$ on SFR is itself strongly dependent on how these properties are measured (see Section \ref{sec:The FMR in observations}). Although a high-mass dependence is undeniable in our Sample T2, it is not present in Sample T1, at least not within the range of masses studied.

Second, the recipes used to model physical processes in \textsc{L-Galaxies}, although up-to-date with current theory, are still rather crude, and this could be affecting the galaxy evolution seen in our model galaxies. For example, metals are assumed to fully mix with the ISM \textit{before} galactic outflows are allowed to drive gas out of the galaxy. This means that it is the subsequent cessation of star formation in low-mass galaxies that is causing the relation between $M_{*}$ and $Z$ in the model, rather than explicitly metal-rich outflows. Additionally, and perhaps relatedly, there is a lack of evolution in the model $M_{*}$-$Z$ relation, contrary to observations.

Despite these two caveats, we believe our interpretation to be viable. On the observational side, the use of many emission line fluxes and SED fitting (as is done for many trusted stellar mass estimations) is likely to be a more robust way of estimating metallicities than using individual emission line ratios (see Appendix A). On the modelling side, the problems outlined above are currently common to models in general. For example, the SPH simulation described by \citet{D11}, which exhibits the same evolution of galaxies along the present-day $M_{*}$-$Z$ relation, also has difficulties reproducing the observed metallicity evolution. Their favoured \textit{vzw} model (which invokes momentum-driven winds and metal-rich outflows at all masses) shows an increase in $Z$ of $\sim 0.15$ dex from $z=2$ to 0 for galaxies of stellar mass $\sim 10^{9.5} \textnormal{M}_{\textnormal{\astrosun}}$. Their fixed, low wind velocity \textit{sw} model shows an evolution of $\sim 0.1$ dex (an amount similar to that seen in \textsc{L-Galaxies}). In comparison, observations suggest an evolution of $\sim 0.5$ dex at the same mass from $z=2.3$ to 0.07 \citep{M08}. Further improvements to the observational determination of gas-phase metallicities, and the ongoing improvements in accurate modelling of galactic evolution, are necessary before significant progress can be made in this area.

\section{Conclusions} \label{sec:Conclusions}
We have shown that the gas-phase metallicities of galaxies are dependent on their star formation rate. This is also true at high masses, where highly star forming galaxies are seen to have higher metallicities -- the opposite trend to that seen at low masses. However, despite this dependence, a projection onto the $M_{*}$-SFR-$Z$ space that combines $M_{*}$ and SFR does little to reduce the scatter in $Z$ compared to the $M_{*}$-$Z$ relation.

We also demonstrate the significance of metallicity derivation methods when assessing the relation between $M_{*}$, SFR and $Z$. Strong line ratio diagnostics, such as R$_{23}$, provide significantly different metallicity estimates to Bayesian techniques which utilise emission line fluxes. These differences are greater for low star forming galaxies. Although we believe that the Bayesian technique used for our Sample T2 provides a more robust measurement of the global gas-phase metallicity in local, high-metallicity, star forming galaxies, it remains unclear to what extent this is so.

In Section \ref{Model Results}, we show that a high-mass SFR-dependence is also present in our model sample. This is due to a turnover in the mass-metallicity relation, caused by a gradual dilution of the gas phase in some galaxies, triggered by a gas-rich merger which shuts down subsequent star formation without impeding further cooling. We have proposed that similarities between these low-sSFR model galaxies and those observed at $z=0$, such as their larger-than-average black hole masses, leave open the possibility that such a process is also driving the SFR-dependence seen at high masses in real galaxies. If this is the case, then physical processes other than mass-dependent outflows must also be playing a part in regulating metallicity. Our model indicates metal-poor galactic infall as a likely candidate.

\section*{Acknowledgments} \label{sec:Acknowledgements}
R. M. Y. would like to thank Jarle Brinchmann, Gabriella De Lucia, Jonny Elliott, Silvia Fabello, Qi Guo, Bruno Henriques, Roberto Maiolino, Filippo Mannucci, Roderik Overzier, Francesco Shankar, Freeke van de Voort, Simon White and Rob Wiersma for invaluable discussions during the undertaking of this work.

\section*{Appendix A} \label{sec:Appendix A}
As mentioned in Section \ref{sec:The M-Z relation, as a function of SFR}, obtaining accurate estimates of the gas-phase metallicity is not an easy process, as different diagnostics can provide very different results. However, it should be emphasised that the difference seen in Fig. \ref{fig:T1T2_abundance_diff} is not simply due to the inherent discrepancies between $T_{e}$, empirical and theoretical metallicity derivation methods. For the range of metallicities covered by our samples, both take their metallicities via theoretical methods. The difference is instead likely due to the use of either emission line ratios or emission line fluxes (with a Bayesian approach) in the analysis.

On this point, it should be noted that theoretically derived strong line ratios are known to suffer from a number of problems, including degeneracies \citep{KD02,KE08}, sensitivity to the ionisation parameter \citep{KD02, E06c}, saturation at high metallicities \citep{KD02, Li06, E06c, KE08} and inconsistency with $T_{e}$ derived metallicities \citep{K03,Br04,G04,St05}. It is not clear to what extent the technique used for Sample T2 suffers from these effects. There have been some concerns over the treatment of secondary nitrogen in the population synthesis models used to produce the Sample T2 metallicity estimates \citep{Li06,Y07}. Early indications show that accounting for this by excluding the [\textsc{Nii}] (and [\textsc{Sii}]) lines from the Bayesian analysis does not change the high-mass SFR-dependence seen. Removing these lines only seems to have a significant effect at low-masses, strengthening the dependence of Z on SFR seen there.\footnote{It should be noted that a larger number of double-peaked likelihood distributions are produced when removing emission lines from the analysis. This can make it more difficult to determine the true metallicity. Further analysis of this new set of metallicity estimates is needed before more concrete statements can be made.}

There are, however, two particular issues affecting the strong line ratio calibrations used for Sample T1 at high metallicity. First, there is the binning of data by \citet{M08} when calibrating the diagnostics used for our Sample T1. A fit to unbinned data would have heavily biased their diagnostics against the lower metallicities crucial for high redshift studies such as theirs, due to the paucity of low-$Z$ galaxies available in their present-day calibration sample. However, this does mean that their fits are less precise at high metallicities, which is important for their application to local samples such as ours. Their R$_{23}$ diagnostic over-predicts the average metallicity slightly compared to the average values obtained from the \citet{KD02} model for $Z\gtrsim 9.0$ (see fig. 5 in the \citet{M08} paper). This raises the metallicity estimated in this regime somewhat, despite the use of the $[\textnormal{N}\textsc{ii}]/\textnormal{H}\alpha$ diagnostic to bring down the final value.

Second, \citet{N06} have already pointed out that their metallicity derivation method -- which uses the same calibration sample as \citet{M08} -- may over estimate the gas metallicity at $Z>9.0$ by a factor of $\sim 0.1$ dex compared to the \citet{T04} method. This is partly due to the bias towards selecting strongly $[\textnormal{O}\textsc{ii}]$ and $[\textnormal{O}\textsc{iii}]$ emitting galaxies in the sample causing the fit to be steeper at high metallicities. When considering the R$_{23}$ diagnostic, the fact that H$\beta$ emission is also weaker in low star formation environments could explain why the discrepancy we see is more significant at low-SFR.

Taking this into account, it is perhaps prudent to suggest that such ratios are not ideal for estimating metallicities for high-$Z$, local samples, where the availability of good spectroscopic data, including all optical emission lines, allows alternative techniques to be utilised.

\renewcommand{\thefigure}{B.\arabic{figure}}
\setcounter{figure}{0}

\section*{Appendix B} \label{sec:Appendix B}
The adaptation made to Sample T2 in Section \ref{sec:Black hole masses} is to include an additional set of 9,275 high-mass ($\geq 10^{10.5} \textnormal{M}_{\textnormal{\astrosun}}$) galaxies to the sample. These are the galaxies removed from the default Sample T2 due to having 1$\sigma$ uncertainties in their stellar mass estimates greater than 0.2 dex. The reason for this larger error seems to be due to large errors in the u-band magnitude measured, which is included in the estimation of stellar mass. This error has therefore likely propagated through to the confidence in the best fitting model during SED fitting, causing these galaxies to have a larger uncertainty in their $M_{*}$ estimate.

\begin{figure}
\centering
\includegraphics[totalheight=0.22\textheight, width=0.36\textwidth]{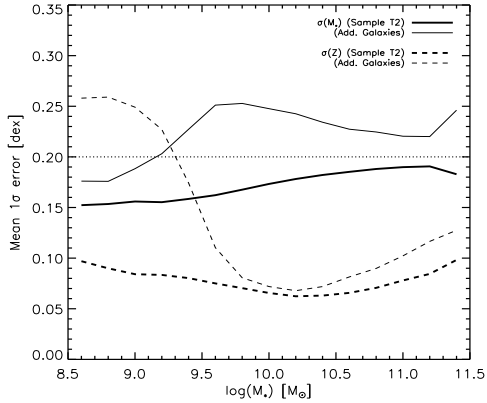}
\caption{The mean 1$\sigma$ errors on the values of $M_{*}$ (solid lines) and $Z$ (dashed lines) provided by the best fitting models for the default Sample T2 (thick lines), and for the additional galaxies described in Appendix B (thin lines). At high masses, the error in $M_{*}$ for the additional galaxies is only slightly greater than for the Sample T2 galaxies. The metallicity errors in the high-mass regime are well below the maximum acceptable error of $\sigma = 0.2$ dex (indicated by the horizontal dotted line).}
\label{fig:T2_MZ_errors}
\end{figure}

Fig. \ref{fig:T2_MZ_errors} shows that, for masses greater than $\sim 10^{10.5} \textnormal{M}_{\textnormal{\astrosun}}$, these additional galaxies have only slightly greater 1$\sigma$ errors on their mass estimates than the default Sample T2 galaxies, and that their metallicity estimates are actually more than good enough for them to remain in the sample. We therefore choose to included these high-mass galaxies for our analysis of black hole mass. Their addition is not a necessary condition, but does provide more low-sSFR galaxies and makes the dichotomy seen in the right panel of Fig. \ref{fig:sSFR-Z_BH} clearer.

\begin{figure}
\centering
\includegraphics[totalheight=0.22\textheight, width=0.36\textwidth]{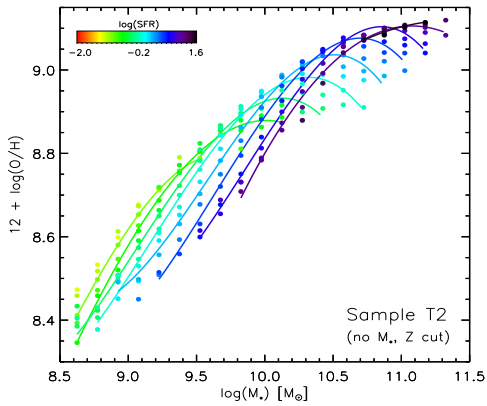}
\caption{The $M_{*}$-$Z$ relation for Sample T2, when including the additional galaxies described in Appendix B. The shape of the overall relation is unchanged, but turnovers in the relation at fixed-SFR are now evident for wide a range of SFRs. This lends favour to the conclusion from our model sample, that processes other than outflows are regulating metallicity at high mass.}
\label{fig:T2_MZR_noMZcut}
\end{figure}

It is interesting to note that when plotting the $M_{*}$-$Z$ relation for Sample T2 including these additional galaxies, a range of SFR-dependent turnovers can be clearly seen at high mass. This is show in Fig. \ref{fig:T2_MZR_noMZcut}. When considering Fig. \ref{fig:sSFR-Z_BH}, Fig. \ref{fig:BHDists} and Fig. \ref{fig:T2_MZR_noMZcut} together, it seems plausible that the low-$Z$, high-$M_{*}$ galaxies with large SMBHs seen in our model sample are also present in the SDSS, and that processes other than outflows are at play in regulating their metallicities.

%The metallicity errors within the high-mass regime are well within the acceptable range of $\sigma < 0.2$ dex (indicated by the dotted line).

\end{document}